\documentclass[aps,pre,10pt,twocolumn,amsmath,amssymb,superscriptaddress,floatfix,showpacs]{revtex4-1}

\usepackage{graphicx}

\usepackage{hyperref}

\renewcommand{\Im}{\mathop\mathrm{Im}\nolimits}

 \DeclareMathOperator{\Tr}{Tr}

\begin{document}

\title{Odd-frequency superconducting states with different types of Meissner response: Problem of coexistence}

\author{Ya.~V.\ Fominov}
\affiliation{L.~D.\ Landau Institute for Theoretical Physics, RAS, 142432 Chernogolovka, Russia}
\affiliation{Moscow Institute of Physics and Technology, 141700 Dolgoprudny, Russia}

\author{Y.\ Tanaka}
\affiliation{Department of Applied Physics, Nagoya University, Nagoya 464-8603, Japan}
\affiliation{Moscow Institute of Physics and Technology, 141700 Dolgoprudny, Russia}

\author{Y.\ Asano}
\affiliation{Department of Applied Physics \& Center for Topological Science and Technology,
Hokkaido University, Sapporo 060-8628, Japan}
\affiliation{Moscow Institute of Physics and Technology, 141700 Dolgoprudny, Russia}

\author{M.\ Eschrig}
\affiliation{Department of Physics, Royal Holloway,
University of London, Egham, Surrey TW20 0EX, United Kingdom}

\date{3 May 2015}

\begin{abstract}
We consider physical properties of a superconductor with a recently proposed type of odd-frequency pairing that exhibits diamagnetic Meissner response (``odd-dia state''). Such a state was suggested in order to address stability issues arising in an odd-frequency superconducting state with paramagnetic Meissner response (``odd-para state''). Assuming the existence of an odd-dia state (due to a proper retarded interaction), we study its coexistence with an odd-para state. The latter is known to be generated as an induced
superconducting component in, e.g., singlet
superconductor/ferromagnet proximity structures or triplet
superconductor/normal metal systems. Calculating the superfluid
density of the mixed odd-para/odd-dia state and the Josephson
current between the odd-para and odd-dia states, we find that
the expressions for the currents in both cases have non-vanishing imaginary contributions and are
therefore unphysical. We show that a realization of the odd-dia state
implies the absence of a Hamiltonian description of the system,
and suggest that there exists no physically realizable perturbation that could give rise to the spontaneous symmetry breaking necessary for an actual realization of the odd-dia superconducting state.
\end{abstract}
\pacs{74.20.Mn, 74.25.N-, 74.20.Rp}
\maketitle
\tableofcontents

\section{Introduction}

Pairing of fermions in superconductors and superfluids can be described by the so-called anomalous Green function $F(1;2)$
(here 1 and 2 denote sets of coordinates for the two fermions). The Pauli principle requires antisymmetry under permutation of two fermions in a Cooper pair, $F(2;1) = -F(1;2)$, leading to the standard classification \cite{Mineev} of superconducting phases:
even parity under exchange of the spatial coordinates of $F$ (\textit{s}-wave, \textit{d}-wave, etc.) must be accompanied by
odd parity under exchange of the spin coordinates (singlet). Similarly,
odd parity in the spatial coordinates (\textit{p}-wave, \textit{f}-wave, etc.) must be accompanied by even parity in the spin coordinates (triplet).
This classification scheme assumes that $F$ stays unchanged under permutation of two imaginary-time coordinates (we assume for definiteness the Matsubara representation).

In 1974, Berezinskii pointed out \cite{Berez} that this classification, implicitly assuming even dependence of $F$ on the imaginary time difference $\tau = \tau_1-\tau_2$, can be doubled if one allows for an odd symmetry with respect to time (or to the Matsubara frequency $\omega$ in the Fourier representation). This opens up a possibility of, e.g., an \textit{s}-wave triplet \cite{Berez,KB} or a \textit{p}-wave singlet \cite{Balatsky92} phase without violating the
Pauli principle or the Fermi-Dirac statistics.
These exotic phases are called odd-frequency (odd-$\omega$) states.
Although there has been much theoretical work on
odd-frequency pairing states (as the principal state, i.e., the state corresponding to a normal-state instability in the pairing interaction channel) in bulk systems
\cite{Emery,Balatsky93,Abrahams,Coleman,Heid,Abrahams95,Martisovots,Zacher,Jarrell,BK,Coleman99,
Vojta,Anders,Fuseya2003,Sakai2004,Hotta,Shigeta1d,ShigetaAF,KFM,Dahal,Yanagi12,Hoshino},
experimental evidence is still lacking.

However, it turns out that the \textit{s}-wave triplet state can be realized as an induced (as opposed to spontaneously) symmetry-broken state, leading to an
induced superconducting component in a rather conventional system consisting of an \textit{s}-wave singlet superconductor (S) and a ferromagnet (F) \cite{BVE}. Ferromagnetism here breaks the symmetry in the spin space and opens up the possibility of generating triplet superconducting correlations from singlet ones \cite{BVE,Tokuyasu1988,Kadigrobov01,Eschrig03,Lofwander05,Eschrig07,Asano2007a,Yokoyama2007,Houzet07,Nazarov07,
Eschrig08,Halterman08,Galaktionov08,Grein09,Beri09,Volkov10,Alidoust10,Trifunovic10,Meng13}.
Alternatively, if the triplet correlations are already present in the system due to, e.g., a principal even-$\omega$ \textit{p}-wave triplet state (TS), then an odd-$\omega$ \textit{s}-wave triplet state can be generated
due to breaking the isotropy in real space \cite{Eschrig07,Tanaka,Tanaka07e}.
This can be achieved, e.g., by contacting the superconductor with a diffusive normal metal (DN) \cite{Tanaka,Tanaka07e}.
Odd-frequency superconductivity in such systems is an emergent inhomogeneous phase appearing in the vicinity of an interface and penetrating into the adjacent materials
due to the proximity effect \cite{Eschrig07,Tanaka07e,Tanuma}.

Induced odd-frequency correlations have also been discussed in connection with surface Andreev bound states in unconventional superconductors \cite{Buchholtz,Hara,Hu,Tanaka95,Kashiwaya,Suzuki}, vortex systems \cite{Yokoyamavortex,Tanuma09,Yokoyama10}, Majorana fermions in topological superconductors \cite{Asano2013,Stanev,Wakatsuki,Balatsky1,Ebisu,Vorontsov,GrMiEs}, and multi-band systems \cite{Balatsky2}.

An unexpected feature of the induced \textit{s}-wave triplet state in diffusive proximity structures is that the resulting local density of state has a zero-energy peak \cite{Tanaka2004,Tanaka,Tanaka2006,Asano2007,Fominov,Linder09}.
For ballistic structures, where both even-$\omega$ (e.g., \textit{p}-wave) and odd-$\omega$ (e.g., \textit{s}-wave) triplet states are generated at the superconductor-ferromagnet interface, a low-energy Andreev bound state band around zero energy was found \cite{Eschrig03,Eschrig09}.
Furthermore,
the sign of the local current response to an external vector potential turns out to be unconventional
(in other words, the local superfluid density is formally negative, $n_S<0$) \cite{Tanaka2005,Asano,Yokoyama2011,Higashitani,Alidoust}.
This would lead to an instability of
such a principal superconducting state in a bulk homogeneous system
\cite{Heid,HeidBazaliy,Coleman,Abrahams95,CoxZawadowski}. On the other hand, the induced odd-$\omega$ state is stabilized in inhomogeneous proximity systems due to presence of a principal conventional superconductor (note at the same time that under special conditions instability in thin-film proximity structures is also possible \cite{Mironov}). Since the conventional ($n_S>0$) superconducting response corresponds to the diamagnetic Meissner effect, we refer to the unconventional ($n_S<0$) state as a (locally) paramagnetic one (odd-$\omega$--para state) \cite{Higashitani97,Barash}. At the same time, in order to avoid confusion, we note that if one studies the Meissner effect in a proximity structure where the surface region has a prevailing odd-$\omega$--para component \cite{BVE,Tanaka}, then this surface region would show an oscillating Meissner effect \cite{Tanaka2005,Yokoyama2011} (since the sign of $n_S$ enters the screening length under the square root) which finally turns to a diamagnetic one in the bulk of the conventional superconductor.

Although surprising at first sight, the odd-$\omega$--para state is predicted on the basis of microscopic models and is undoubtedly realized as an induced superconducting state in real proximity structures \cite{BVE,Tokuyasu1988,Kadigrobov01,Eschrig03,Lofwander05,Eschrig07,Asano2007a,Yokoyama2007,Houzet07,Nazarov07,
Eschrig08,Halterman08,Galaktionov08,Grein09,Beri09,Volkov10,Alidoust10,Trifunovic10,Meng13,Tanaka,Tanaka07e}. There are numerous experimental results consistent with  manifestations of this state \cite{Keizer,Petrashov06,Anwar,Robinson10,Khaire10,Sprungmann10,Lofwander10,Gu10,Boden,Kalcheim12,Visani12,Yates13,
Piano,HublerWolf,Flokstra}.
An interesting feature of the odd-$\omega$--para state in proximity structures is that it can be spatially separated from other superconducting components arising due to the proximity effect \cite{BVE,Tokuyasu1988,Kadigrobov01,Eschrig03,Lofwander05,Eschrig07,Asano2007a,Yokoyama2007,Houzet07,Nazarov07,
Eschrig08,Halterman08,Galaktionov08,Grein09,Beri09,Volkov10,Alidoust10,Trifunovic10,Meng13,Tanaka,Tanaka07e}. For example, in S/F junctions the so-called long-range proximity effect arises \cite{BVE}, meaning that not too close to the interface the only surviving superconducting component in the F layer has odd-$\omega$--para properties. So, this superconducting state is indeed realized (as the only superconducting component) in certain spatial regions.

Recently, a further extension of the classification of superconducting phases was proposed in Refs.~\cite{Solenov,Kusunose}. The proposed state is also odd-$\omega$, but at the same time it possesses a conventional diamagnetic Meissner response (odd-$\omega$--dia, $n_S>0$), and thus is supposed to solve the problem of instability for a bulk material. If so, this state could be realized as a homogeneous principal superconducting phase. Being more conventional at first glance, this proposal
relies on an essential assumption that a proper retarded interaction leading to this state can exist in a real material.

Motivated by this proposal, we study physical properties of the odd-$\omega$--dia state (assuming it is actually realized). It turns out that while the new state seems consistent by itself, coexistence of the two different odd-$\omega$ states (para and dia) leads to unphysical consequences. Below, we describe the inconsistencies arising from the assumption of such coexistence and propose our view on a possible resolution of the contradiction.

The paper is organized as follows. In Sec.~\ref{sec:pathint}, we review the path-integral approach of Refs.~\cite{Solenov,Kusunose} in order to establish the notations and underline the features that are essential for further discussion. In Sec.~\ref{sec:bulk}, we consider possibility of coexistence of the odd-$\omega$--para and odd-$\omega$--dia states in the same spatial region. In Sec.~\ref{sec:Josephson}, we consider the possibility of their coexistence in the Josephson junction. Our results are then discussed in Sec.~\ref{sec:discussion} and summarized in Sec.~\ref{sec:conclusions}.

\section{Path-integral formulation}
\label{sec:pathint}

In this section, we develop the path-integral formalism for describing superconductivity in both homogeneous and inhomogeneous states with general interactions. We then proceed to discuss a specific type of interactions, following the path-integral approach of Refs.~\cite{Solenov,Kusunose} for describing the homogeneous principal odd-$\omega$--dia state. Reference~\cite{Solenov} treated spinless fermions, while the general case of arbitrary pairing for fermions with spin was considered in the Appendix of Ref.~\cite{Kusunose}. We briefly outline the main points of the derivation, following Ref.~\cite{Kusunose}, with intention to set up the general framework necessary for our later discussion of coexistence between the odd-$\omega$--dia and odd-$\omega$--para states.
We pay special attention to differences between description of the odd-$\omega$--dia and --para states.

The reader less interested in the technical details of a rigorous treatment can skip the following subsection which offers a general treatment, and proceed directly to the next subsection that presents a simplified treatment valid for a more restricted set of interactions.

We adopt the following definition for the complex conjugation of Grassmann variables:
$\psi^{**} = \psi$, $(\psi_1 \psi_2)^* = \psi_2^* \psi_1^*$. The temperature is denoted as $T$, and we put $\hbar=k_\mathrm{B} =1$ throughout the paper.

\subsection{General treatment}
\label{sec:general}

\subsubsection{Interaction}
We consider a general interaction
\begin{equation} \label{tildeVgeneral}
\tilde V_{\alpha\beta;\gamma\delta}(\mathbf r_1\tau_1,\mathbf r_2\tau_2;\mathbf r_3\tau_3, \mathbf r_4\tau_4)
\end{equation}
as matrix $\tilde V(12;34)$ with compound index $12$ denoting the set of variables $\mathbf{r}_1\tau_1\alpha,\mathbf r_2\tau_2\beta$ and compound index $34$ denoting the set of variables $\mathbf r_3\tau_3\gamma,\mathbf r_4\tau_4\delta $.
Here, $\alpha, \beta,\gamma, \delta$ are spin indices, while the imaginary times (we use the Matsubara representation) are denoted $\tau_1, \ldots, \tau_4$, and the spatial coordinates are
$\mathbf r_1,\ldots,\mathbf r_4$.
We use a brief notation for integration
\begin{equation}
\int_1 \equiv \int_0^{1/T} d\tau_1 \int d\mathbf r_1 \sum_\alpha , \quad \int_{12}\equiv \int_1 \int_2, \quad \mbox{etc}.
\end{equation}
We will use the singular value decomposition for the interaction matrix defined as above.
Accordingly, the interaction matrix can be expanded into the singular vectors
\begin{equation}
\tilde V(12;34) = \int_{\lambda } \zeta^{\lambda }(12) \;  V_\lambda \; \eta^{\lambda }(34)^* ,
\end{equation}
where $\lambda$ labels a
complete orthonormal set of left singular vectors $\zeta^\lambda (12)$,
\begin{subequations}
\begin{align}
\int_{12} \zeta^\lambda(12)^* \zeta^{\lambda'}(12) &= \delta_{\lambda\lambda'} ,
\\
\int_{\lambda} \zeta^\lambda(12) \zeta^{\lambda}(34)^* &= \delta_{13} \delta_{24},
\end{align}
\end{subequations}
as well as correspondingly for the right singular vectors $\eta^\lambda(34)$.
We use a brief delta-functional notation, e.g., $\delta_{13} \equiv \delta(\mathbf r_1-\mathbf r_3) \delta(\tau_1-\tau_3) \delta_{\alpha\gamma}$.

The singular values $V_\lambda$ are real and can be assumed all non-positive (we slightly modify the convention for the usual definition in terms of non-negative singular values). We will also use the pseudo-inverse of this matrix, defined as
\begin{equation}
\tilde V^{-p}(12;34) = \int_{\lambda } \eta^{\lambda }(12) \; V^{-p}_\lambda \; \zeta^{\lambda }(34)^*,
\end{equation}
where $V^{-p}_\lambda $ is obtained from $V_\lambda$ by replacing all non-zero singular values by their inverses, leaving all zero singular values untouched.

Below, for brevity we employ the matrix product notation
\begin{subequations}
\begin{align}
[AB](12;34) &\equiv \int_{56} A(12;56) B(56;34),
\\
[aA](34) &\equiv \int_{12} a(12) A(12;34),
\\
[Aa](12) &\equiv \int_{34} A(12;34) a(34),
\end{align}
\end{subequations}
for dealing with four-point functions ($A$, $B$) and two-point functions ($a$).

Then
$P_1(12;34)\equiv [\tilde V\tilde V^{-p}](12;34)$
and
$P_2(12;34)\equiv [\tilde V^{-p}\tilde V](12;34)$
turn out to be Hermitian projector matrices (not necessarily identity matrices), and $[\tilde V^{-p}\tilde V\tilde V^{-p}]=\tilde V^{-p}$ as well as $[\tilde V\tilde V^{-p}\tilde V]=\tilde V$.

\subsubsection{Partition function and action}
The partition function $\mathcal Z$ is written in the path-integral formulation with the help of Grassmann fields $\psi(1)$ and $\psi^*(1)$:
\begin{gather}
\mathcal Z = \int D\psi^* D\psi \exp(- \mathcal S_0 - \mathcal S_\mathrm{int}),
\label{Z} \\
\mathcal S_0 = \int_1 \psi^*(1) \left( \partial_{\tau_1} + \xi \right) \psi(1),
\label{S0fermwithspin} \\
\mathcal S_\mathrm{int} = \frac{1}{2} \int_{1234} \rho^*(12) \; \tilde V(12;34) \; \rho(34),
\label{Sintspinsgen}
\end{gather}
where $\xi = -\partial_{\mathbf r_1}^2 /2m - \mu$ is the kinetic energy counted from the chemical potential,
and notations
\begin{equation}
\rho(12) = \psi(1) \psi (2),  \quad
\rho^*(12) = \psi^*(2) \psi^*(1)
\end{equation}
are introduced for brevity for the pair-density field and its complex conjugate.

The $\mathcal S_0$ action describes free particles, while $\mathcal S_\mathrm{int}$ describes spin-dependent interaction.
Requiring $\mathcal Z= \mathcal Z^*$
and changing the integration variables (Grassmann fields) into
$\eta(\mathbf r,\tau,\alpha) = \psi(\mathbf r,1/T-\tau,\alpha)$,
$\eta^*(\mathbf r,\tau,\alpha) = \psi^*(\mathbf r,1/T-\tau,\alpha)$,
we obtain the condition
\begin{equation}
\tilde V(12;34) = \tilde V^*(\bar 3\bar 4;\bar 1\bar 2),
\end{equation}
with $\bar 1\equiv (\mathbf r_1,1/T - \tau_1,\alpha)$, etc. This is equivalent to
\begin{equation} \label{Vfermexchgen}
V_\lambda \int_{12} \eta^\lambda(12)^* \zeta^{\lambda'}(\bar 1\bar 2) = V_{\lambda'} \int_{12} \zeta^\lambda(12)^* \eta^{\lambda'}(\bar 1 \bar 2) .
\end{equation}
At the same time, $\tilde V$ fulfills the relations
\begin{equation} \label{Vfermexchgen2}
\tilde V(12;34) = \tilde V(21;43)=-\tilde V(21;34)=-\tilde V(12;43),
\end{equation}
conditions that follow directly from exchanging integration variables in $\mathcal S_\mathrm{int}$.
These conditions are equivalent to the relations
\begin{equation}
\zeta^\lambda(12)=-\zeta^\lambda(21), \quad \eta^\lambda(12)=-\eta^\lambda(21),
\quad \mbox{$\lambda\notin \Lambda_0$},
\end{equation}
where $\Lambda_0$ is the set of all $\lambda $ for which $V_\lambda=0$.
With the definitions
\begin{equation}
\rho_\lambda = \int_{12} \eta^\lambda(12)^* \rho(12), \quad
\rho^+_\lambda = \int_{12} \rho^*(12) \zeta^\lambda(12),
\end{equation}
we obtain finally
\begin{equation}
\mathcal S_\mathrm{int} = \frac{1}{2} \int_{\lambda} \rho^+_{\lambda } \; V_\lambda  \; \rho_\lambda .
\end{equation}

\subsubsection{Mean-field approximation}
Performing a Hubbard-Stratonovich transformation amounts to multiplying the partition function by a finite constant
\begin{equation} \label{HSgen}
\int D\overline\Delta D\Delta
\exp\biggl\{ \frac{1}{2} \int_{1234}
\overline\Delta (12)
\tilde V^{-p}(12;34)
\Delta(34)
\biggr\},
\end{equation}
and subsequently shifting the integration fields as
$\Delta \mapsto \Delta + [\tilde V\rho]$ and $\overline\Delta \mapsto \overline\Delta + [\rho^* \tilde V]$,
using $[(\overline\Delta +\rho^* \tilde V) \tilde V^{-p} (\Delta + \tilde V\rho )]= [\overline\Delta \tilde V^{-p} \Delta ]+
[\rho^* P_1 \Delta] + [\overline\Delta P_2 \rho ]
+ [\rho^* \tilde V \rho ]$.
As a result, the four-fermionic term $\mathcal S_\mathrm{int}$ is decoupled:
\begin{equation} \label{dec}
\exp(-\mathcal S_\mathrm{int}) \mapsto \int D\overline\Delta D\Delta \exp(-\mathcal S_\mathrm{aux}-\mathcal S_\Delta),
\end{equation}
where
\begin{align}
\mathcal S_\mathrm{aux}= -\frac{1}{2} \int_{12}
& \Big( \rho^*(12) [P_1\Delta](12) +  [\overline\Delta P_2](12) \rho (12) \Big), \nonumber
\\
\mathcal S_\Delta= -\frac{1}{2} \int_{1234}
& [\overline\Delta P_2](12) \tilde V^{-p}(12;34) [P_1\Delta](34) .
\end{align}
Defining the fields
\begin{equation}
\Delta_\lambda = \int_{12} \zeta^\lambda(12)^* \Delta(12), \quad
\overline\Delta_\lambda = \int_{12} \overline\Delta(12) \eta^\lambda(12)
\end{equation}
for all $\lambda \notin \Lambda_0$,
and setting $\Delta_\lambda=\overline\Delta_\lambda=0$ for all $\lambda \in \Lambda_0$, we obtain
\begin{align}
\mathcal S_\mathrm{aux} &= -\frac{1}{2} \int_\lambda
\left( \rho^+_\lambda \Delta_\lambda + \overline\Delta_\lambda  \rho_\lambda \right),
\\
\mathcal S_\Delta &= -\frac{1}{2} \int_\lambda
\left( \overline\Delta_\lambda V^{-p}_\lambda \Delta_\lambda \right).
\end{align}
We require $\overline\Delta_\lambda = \Delta^*_\lambda $ for ensuring convergence in the Hubbard-Stratonovich transformation.

\subsubsection{Saddle-point solution}
Now, we focus on finding the saddle-point solution.
Instead of integrating over $\overline\Delta$  and $\Delta$, we take a trial path with respect to which we will vary the action.
Then, we can define the mean-field free-energy functional:
\begin{multline}
\mathcal F_\mathrm{MF} [ \overline\Delta, \Delta ] = - T \ln \mathcal Z_\mathrm{MF}
\\
= - T \ln \int D\psi^* D\psi \exp\left( -\mathcal S_\mathrm{MF}[\psi^*,\psi,\overline\Delta,\Delta] \right),
\end{multline}
where $\mathcal S_\mathrm{MF} = \mathcal S_0 +\mathcal S_\mathrm{aux} +\mathcal S_\Delta$ is the mean-field action in which $\overline\Delta$ and $\Delta$ are replaced by the trial path.
To find the saddle point, we should minimize $\mathcal F_\mathrm{MF}$ with respect to the trial path.
As a result, we obtain the self-consistency equations for the mean-field pair potential $\Delta_\mathrm{MF}$ (also called the gap function),
\begin{subequations} \label{matricesSgen}
\begin{align}
\Delta_\mathrm{MF}(12)\equiv [P_1\Delta](12) &= -\int_{34}\tilde V(12;34) F(34),
\\
\overline\Delta_\mathrm{MF}(34)\equiv
[\overline\Delta P_2](34) &= -\int_{12} F^+(21) \tilde V(12;34),
\end{align}
\end{subequations}
where we have introduced the anomalous averages (anomalous Green functions),
\begin{equation} \label{FbarFgen}
F(12) = \bigl< \rho(12) \bigr>_\mathrm{MF},
\quad
F^+(21) = \bigl< \rho^* (12) \bigr>_\mathrm{MF},
\end{equation}
with the help of the definition of the mean-field averaging:
\begin{equation} \label{MFavgen}
\left< \dots \right>_\mathrm{MF} = \frac{\int D\psi^* D\psi\; (\dots) e^{-\mathcal S_\mathrm{MF} [\psi^*,\psi,\overline\Delta,\Delta]}}{\int D\psi^* D\psi\; e^{-\mathcal S_\mathrm{MF} [\psi^*,\psi,\overline\Delta,\Delta]}}.
\end{equation}
The notations $F$ and $F^+$ for the two anomalous averages are standard in the theory of superconductivity, and the $+$ superscript should not be confused with the Hermitian conjugation $\dagger$.
From the relations
\begin{equation}
F(12)=-F(21), \quad F^+(12)=-F^+(21),
\end{equation}
following directly from the properties of Grassmann variables, and from Eqs.\ (\ref{Vfermexchgen2}),
one obtains the corresponding relations
\begin{equation}
\Delta_\mathrm{MF}(12)=-\Delta_\mathrm{MF}(21), \quad
\overline\Delta_\mathrm{MF}(12)=-\overline\Delta_\mathrm{MF}(21).
\end{equation}
Finally, with the definitions
\begin{equation}
F_\lambda = \int_{12} \eta^\lambda(12)^* F(12) , \quad
F^+_\lambda = \int_{12} F^+(21) \zeta^\lambda(12)
\end{equation}
(note the difference in the definitions for $F_\lambda $ and $\Delta_\lambda $, and for $F^+_\lambda $ and $\overline\Delta_\lambda $, correspondingly),
we obtain
\begin{equation}
\Delta_\lambda = -V_\lambda F_\lambda, \quad
\Delta^*_\lambda = -V_\lambda F^+_\lambda .
\end{equation}
From these relations it follows that, for all components $F_\lambda$, $F^+_\lambda $ for which $\lambda \notin \Lambda_0$, the symmetry relation $F^+_\lambda = F^*_\lambda $ holds (remember that $V_\lambda $ is real). For all $F_\lambda$, $F^+_\lambda $ with $\lambda \in \Lambda_0$ nothing follows from these equations (as then $V_\lambda=\Delta_\lambda=\Delta^*_\lambda=0$).

\subsection{Simplified treatment}
\subsubsection{Interaction}

In order to make our consideration more transparent and facilitate comparison with Refs.~\cite{Solenov,Kusunose}, we now assume that the interaction is homogeneous with respect to spatial coordinates and time. For that we slightly change notations, indicating the spin indices explicitly, while the spatial coordinates and imaginary times are gathered in 4-vectors denoted as
$1\equiv x_1 = (\mathbf r_1,\tau_1)$ and $2\equiv x_2 = (\mathbf r_2,\tau_2)$.
The general interaction of Sec.~\ref{sec:general} [Eq.\ (\ref{tildeVgeneral})] now takes the form $V_{\alpha\beta;\gamma\delta}(1-2) \delta(1-3) \delta(2-4)$, where we denote $\delta(1-3)\equiv \delta(\mathbf r_1-\mathbf r_3) \delta(\tau_1-\tau_3)$, etc.

At the same time, we keep general form with respect to interactions in both the singlet and triplet channels (so that later we can consider specific cases on the basis of general equations). So, we assume interaction of the form
\begin{multline}
V_{\alpha \beta;\gamma\delta}(1-2) =
V_s(1-2) \frac{(i\sigma_2)_{\alpha\beta} (i\sigma_2)^*_{\gamma\delta}}{2}
\\
+\sum_{j=1}^3 V_t^{(j)}(1-2) \frac{(i\sigma_j \sigma_2)_{\alpha\beta} (i\sigma_j \sigma_2)^*_{\gamma\delta}}2.
\end{multline}

The singlet interaction $V_s(1-2)$ and the three components of the triplet interaction $V_t^{(j)}(1-2)$ are assumed to be either negative definite or zero \cite{BCScomm}.
For this case, the matrix inverse of the interaction matrix with two compound spin indices is defined as
\begin{multline}
[V^{-1}(1-2)]_{\alpha \beta;\gamma\delta} =
V_s(1-2)^{-1} \frac{(i\sigma_2)_{\alpha\beta} (i\sigma_2)^*_{\gamma\delta}}{2}
\\
+\sum_{j=1}^3 V_t^{(j)}(1-2)^{-1} \frac{(i\sigma_j \sigma_2)_{\alpha\beta} (i\sigma_j \sigma_2)^*_{\gamma\delta}}2,
\end{multline}
with $V(1-2)^{-1} \equiv 1/V(1-2)$ for nonzero components and zero otherwise.

\subsubsection{Partition function and action}
The partition function $\mathcal Z$ is written in the path-integral formulation with the help of
Grassmann fields $\psi_{\alpha}(1)$ and $\psi_{\alpha}^*(1)$:
\begin{gather}
\mathcal Z = \int D\psi_\uparrow^* D\psi_\downarrow^* D\psi_\uparrow D\psi_\downarrow \exp(- \mathcal S_0 - \mathcal S_\mathrm{int}),
\label{Z_s} \\
\mathcal S_0 = \int_1 \psi_\alpha^*(1) \left( \partial_{\tau_1} + \xi \right) \psi_\alpha(1),
\label{S0fermwithspin_s} \\
\mathcal S_\mathrm{int} = \frac{1}{2} \int_{12} V_{\alpha\beta;\gamma\delta}(1-2) \rho_{\alpha\beta}^*(1,2) \rho_{\gamma\delta}(1,2),
\label{Sintspins}
\end{gather}
where $\xi = -\partial_{\mathbf r_1}^2 /2m - \mu$ is the kinetic energy counted from the chemical potential,
summation over repeated spin indices ($\alpha, \beta, \gamma,\delta$) is implied,
and notations
\begin{equation}
\rho_{\alpha\beta}(1,2) = \psi_\alpha(1) \psi_\beta(2), \quad
\rho_{\alpha\beta}^*(1,2) = \psi_\beta^*(2) \psi_\alpha^*(1)
\end{equation}
are introduced for brevity for the pair-density field and its complex conjugate.
The brief notation for integration is
\begin{equation}
\int_1 \equiv \int_0^{1/T} d\tau_1 \int d\mathbf r_1, \quad
\int_{12} \equiv \int_1\int_2, \quad \mbox{etc.}
\end{equation}

The $\mathcal S_0$ action describes free particles, while $\mathcal S_\mathrm{int}$ describes spin-dependent interaction.
Requiring $\mathcal Z= \mathcal Z^*$
and changing the integration variables (Grassmann fields) into
$\eta_\alpha(x) = \psi_\alpha(-x)$,
$\eta_\alpha^*(x) = \psi_\alpha^*(-x)$,
we obtain the condition
\begin{equation}
V_{\alpha\beta;\gamma\delta}(1-2) = V_{\gamma\delta;\alpha\beta}^*(2-1) .
\end{equation}
At the same time, $V$ fulfills the relation
\begin{equation} \label{Vfermexch}
V_{\alpha\beta;\gamma\delta}(1-2) = V_{\beta\alpha;\delta\gamma}(2-1),
\end{equation}
a condition that follows directly from exchanging integration variables in $\mathcal S_\mathrm{int}$.
These two symmetries lead to the requirements that the singlet and triplet interactions, $V_s(1-2)$ and $V_t^{(j)}(1-2)$, are real and even:
\begin{equation} \label{V1m2}
V_s(1-2) = V_s(2-1), \qquad V_t^{(j)}(1-2)=V_t^{(j)}(2-1).
\end{equation}

\subsubsection{Mean-field approximation}
Performing a Hubbard-Stratonovich transformation amounts to multiplying the partition function by a finite constant
\begin{multline} \label{HS}
\int D\Delta^* D\Delta
\exp\biggl\{ \frac{1}{2} \int_{12} \left[ V^{-1}(1-2) \right]_{\alpha\beta;\gamma\delta}
\\
\times \Delta_{\alpha\beta}^* (1,2) \Delta_{\gamma\delta}(1,2) \biggr\},
\end{multline}
and subsequently shifting the integration field as $\Delta_{\alpha\beta}(1,2) \mapsto
\Delta_{\alpha\beta}(1,2) + V_{\alpha\beta;\gamma\delta}(1-2) \rho_{\gamma\delta}(1,2)$.

As a result, the four-fermionic term $\mathcal S_\mathrm{int}$ is decoupled,
\begin{equation}
\exp(-\mathcal S_\mathrm{int}) \mapsto \int D\Delta^* D\Delta \exp(-\mathcal S_\mathrm{aux}-\mathcal S_\Delta),
\end{equation}
with
\begin{align}
&\mathcal S_\mathrm{aux}= -\frac{1}{2} \int_{12}
\left[ \Delta_{\alpha\beta}(1,2) \rho_{\alpha\beta}^*(1,2) + \Delta_{\alpha\beta}^*(1,2) \rho_{\alpha\beta}(1,2) \right],
\\
&\mathcal S_\Delta= -\frac{1}{2} \int_{12}
\left[ V^{-1}(1-2) \right]_{\alpha\beta;\gamma\delta} \Delta_{\alpha\beta}^*(1,2) \Delta_{\gamma\delta}(1,2).
\end{align}
Restricting the Hubbard-Stratonovich transformation to negative definite $V(1-2)$
means that the quadratic form in the exponent of Eq.\ (\ref{HS}) is negative definite, ensuring that the integration in Eq.\ (\ref{HS}) is convergent.

\subsubsection{Saddle-point solution}
Now, we focus on finding the saddle-point solution.
Instead of integrating over $\Delta^*$ and $\Delta$, we take a trial path with respect to which we will vary the action.
Then, we can define the mean-field free-energy functional:
\begin{multline}
\mathcal F_\mathrm{MF} [ \Delta^*, \Delta ] = - T \ln \mathcal Z_\mathrm{MF}
\\
= - T \ln \int D\psi^* D\psi \exp\left( -\mathcal S_\mathrm{MF}[\psi^*,\psi,\Delta^*,\Delta] \right),
\end{multline}
where $\mathcal S_\mathrm{MF} = \mathcal S_0 +\mathcal S_\mathrm{aux} +\mathcal S_\Delta$ is the mean-field action in which $\Delta^*(1,2)$ and $\Delta(1,2)$ are replaced by the trial path.
Further below we will treat the homogeneous case, for which the saddle-point solution does not depend on the center-of-mass coordinate $(x_1+x_2)/2$; for this case one chooses the trial path $\Delta^*(x)$, $\Delta(x)$ depending only on the relative coordinate $x= x_1-x_2$.

To find the saddle point, we should minimize $\mathcal F_\mathrm{MF}$ with respect to the trial path.
As a result, we obtain the self-consistency equations for the mean-field pair potential $\Delta$ (also called the gap function):
\begin{subequations} \label{matricesS}
\begin{align}
\Delta_{\alpha\beta}(1,2) &= -V_{\alpha\beta;\gamma\delta}(1-2) F_{\gamma\delta}(1,2),
\label{matricesSa}
\\
\Delta_{\alpha\beta}^*(1,2) &= -V_{\beta\alpha;\delta\gamma}^*(2-1) F^+_{\delta\gamma}(2,1),
\end{align}
\end{subequations}
where we have introduced the anomalous averages (anomalous Green functions),
\begin{subequations} \label{FbarF}
\begin{align}
F_{\alpha\beta} (1,2)
&= \bigl< \psi_\alpha (1) \psi_\beta (2) \bigr>_\mathrm{MF},
\label{F} \\
F^+_{\alpha\beta} (1,2)
&= \bigl< \psi_\alpha^* (1) \psi_\beta^* (2) \bigr>_\mathrm{MF},
\label{barF}
\end{align}
\end{subequations}
with the help of the definition of the mean-field averaging:
\begin{equation} \label{MFav}
\left< \dots \right>_\mathrm{MF} = \frac{\int D\psi^* D\psi\; (\dots) e^{-\mathcal S_\mathrm{MF} [\psi^*,\psi,\Delta^*,\Delta]}}{\int D\psi^* D\psi\; e^{-\mathcal S_\mathrm{MF} [\psi^*,\psi,\Delta^*,\Delta]}}.
\end{equation}
The notations $F$ and $F^+$ for the two anomalous averages are standard in the theory of superconductivity, and the $+$ superscript should not be confused with the Hermitian conjugation $\dagger$.
Grassmann variables ensure fermionic antisymmetry of the anomalous averages:
\begin{equation} \label{Fantisym}
F_{\alpha\beta}(1,2) = - F_{\beta\alpha}(2,1),
\quad
F^+_{\alpha\beta}(1,2) = - F^+_{\beta\alpha}(2,1).
\end{equation}
Together with the symmetry of the interaction, Eq.\ (\ref{Vfermexch}), this property translates into fermionic antisymmetry of the pair potential,
\begin{equation} \label{Deltaantisym}
\Delta_{\alpha\beta}(1,2) = - \Delta_{\beta\alpha}(2,1).
\end{equation}
Introducing for notational simplicity
\begin{equation} \label{Delta12}
\Delta^+_{\alpha\beta}(1,2) \equiv \Delta^*_{\beta\alpha}(2,1),
\end{equation}
we can write the fermionic part of the action in the form
\begin{align}
\mathcal S_0 +\mathcal S_\mathrm{aux}
&=
\frac 12 \int_{12}
\left( \psi_\alpha^*(1), \psi_\alpha(1) \right) \hat M_{\alpha\beta} (1,2)
\begin{pmatrix}
\psi_\beta(2) \\
\psi_\beta^*(2)
\end{pmatrix},
\label{Squad}
\\
\hat M_{\alpha\beta} (1,2)
&=
\begin{pmatrix}
\delta(1-2) \delta_{\alpha\beta} (\partial_\tau+\xi) &
\Delta_{\alpha\beta}(1,2)
\\
\Delta^+_{\alpha\beta}(1,2) &
\delta(1-2) \delta_{\alpha\beta} (\partial_\tau-\xi)
\end{pmatrix}.
\label{M}
\end{align}
The $\hat M$ matrix is written explicitly in the particle-hole space, and each its element is a matrix in the spin space.

In addition to the Gor'kov (anomalous) Green functions (\ref{FbarF}), we also introduce the standard Green functions:
\begin{align}
G_{\alpha\beta}(1,2) &= - \bigl< \psi_\alpha(1) \psi_\beta^*(2) \bigr>_\mathrm{MF},
\\
G'_{\alpha\beta}(1,2) &= - \bigl< \psi_\alpha^*(1) \psi_\beta(2) \bigr>_\mathrm{MF} = -G_{\beta\alpha}(2,1).
\end{align}
The quadratic form of the action (\ref{Squad}) implies that the Green functions are expressed in terms of the $\hat M$ matrix \cite{Kusunose}:
\begin{equation} \label{Minverse}
\hat M_{\alpha\beta}(1,2)
=
\begin{pmatrix}
-G_{\alpha\beta}(1,2) & F_{\alpha\beta}(1,2) \\
F^+_{\alpha\beta}(1,2) & -G'_{\alpha\beta}(1,2)
\end{pmatrix}^{-1}.
\end{equation}

\subsection{Symmetry relations}

In this subsection we discuss a relation between the two types of the anomalous averages, Eqs.\ (\ref{FbarF}), for simplicity focusing mainly on the homogeneous case, when the Green functions depend on the difference of the coordinates, $(1-2)$.
In the homogeneous case, Fourier-transformed functions $A(1,2)$ [see Appendix~\ref{sec:app:Fourier}] take the form $A(k,k') = A(k) \delta\left( [k-k']/2\pi \right)$.

Similarly to the procedure discussed in Ref.~\cite{Solenov} (generalizing it to the case of fermions with spin), we can obtain a relation between $F$ and $F^+$ directly from definitions
(\ref{FbarFgen}) and
(\ref{FbarF}), applying complex conjugation to one of the anomalous averages. This procedure is nontrivial because in the path-integral formulation with averaging defined according to Eqs.\ (\ref{MFavgen}) and (\ref{MFav}), we have $\mathcal S_\mathrm{MF} \neq \mathcal S_\mathrm{MF}^*$ due to the $\partial_\tau$ term in the action $\mathcal S_\mathrm{MF}$, and then $\left< A \right>_\mathrm{MF}^*$ is not necessarily equal to $\left< A^* \right>_\mathrm{MF}$ in the general case. In order to relate $F^*$ to $F^+$, we have to take into account fermionic antiperiodicity, $\psi_\alpha(\tau+\beta) = -\psi_\alpha(\tau)$, change $\tau\mapsto -\tau$, and define new variables of the path integration $\eta$ depending on the symmetry of $\Delta$: for the case of $\Delta(\tau) = \Delta(-\tau)$, we define $\eta_\alpha(\mathbf r,\tau) = \psi_\alpha(\mathbf r,-\tau)$, $\eta_\alpha^*(\mathbf r,\tau) = \psi_\alpha^*(\mathbf r,-\tau)$, while for the case of $\Delta(\tau) = -\Delta(-\tau)$, we define $\eta_\alpha(\mathbf r,\tau) = i\psi_\alpha(\mathbf r,-\tau)$, $\eta_\alpha^*(\mathbf r,\tau) = -i\psi_\alpha^*(\mathbf r,-\tau)$ \cite{Solenov}. The relation then depends on the symmetry of $\Delta(\tau)$ or $\Delta(\omega)$. We find
\begin{equation} \label{FF+hom}
F_{\alpha\beta}^+(\mathbf k,\omega) = s_\Delta F_{\beta\alpha}^*(\mathbf k,-\omega),
\end{equation}
where $s_\Delta=\pm 1$ for the even-/odd-$\omega$ dependence of $\Delta$. The type of Meissner response is eventually determined by the relative sign between $F_{\alpha\beta}^+(k)$ and $F_{\beta\alpha}^*(k)$, so it depends on both $s_\Delta$ and the frequency symmetry of the anomalous averages.

In the model of Refs.~\cite{Solenov,Kusunose}, it was implicitly assumed that the frequency symmetry of $\Delta$ directly determines (coincides with) the frequency symmetry of $F$ and $F^+$.
In this case, $F_{\alpha\beta}(\mathbf k,-\omega) = s_\Delta F_{\alpha\beta}(\mathbf k,\omega)$, and then Eq.\ (\ref{FF+hom}) immediately yields $F_{\alpha\beta}^+(k) = F_{\beta\alpha}^*(k)$ (both in the even-$\omega$ and odd-$\omega$ cases). This relation corresponds to the conventional diamagnetic Meissner response, therefore the authors of Refs.~\cite{Solenov,Kusunose} concluded that the odd-$\omega$ superconducting state can be realized as a bulk state (if a proper interaction leading to odd-$\omega$ pair potential really exists).

On the other hand, relation (\ref{FF+hom}) is more general and remains valid, for example, if the Zeeman term (exchange energy) is added to the action determined by Eqs.\ (\ref{S0fermwithspin_s}) and (\ref{Sintspins}). This term breaks the symmetry in the spin space and leads to appearance of superconducting components (anomalous averages) with the symmetry differing from that of the interaction $V$ and the pair potential $\Delta$. The simplest example is the ``superconducting ferromagnet'' considered in Appendix~\ref{sec:app:sF}. Since the (instantaneous) interaction in this model is of conventional BCS type \cite{BCScomm}, the pair potential is defined at coinciding imaginary times so that instead of $\Delta(\tau)$ we only have nonzero value $\Delta(0)$. This situation is only compatible with the symmetry class having even-$\tau$ dependence, for which the case of $s_\Delta=1$ is realized. However, the anomalous averages are mixtures of even-$\omega$ singlet and odd-$\omega$ triplet components [see Eqs.\ (\ref{F(k)})--(\ref{F_t})]. So, while $\Delta$ is even-$\omega$, $F$ and $F^+$ do not have definite frequency symmetry.

Similar situation is realized in S/F \cite{Tokuyasu1988,BVE,Kadigrobov01,Eschrig03,Lofwander05,Eschrig07,Asano2007a,Yokoyama2007,Houzet07,Nazarov07,
Eschrig08,Halterman08,Galaktionov08,Grein09,Beri09,Volkov10,Alidoust10,Trifunovic10,Meng13}
and TS/DN \cite{Eschrig07,Tanaka,Tanaka07e,Higashitani09} proximity structures. Generally, this type of odd-$\omega$ states does not correspond to the symmetry of the underlying effective electron-electron attraction, and is induced due to symmetry breaking either in spin or coordinate space. At the same time, $\Delta$ belongs to the even-$\omega$ class so that relation (\ref{FF+hom}) with $s_\Delta=1$ is realized. This relation is valid for all components of the anomalous averages. For odd-$\omega$ components, it yields $F_{\alpha\beta}^+(k) = -F_{\beta\alpha}^*(k)$, which corresponds to the paramagnetic Meissner response.

Note that relation (\ref{FF+hom}) was obtained directly from the definitions of the anomalous averages. One may suggest that an alternative way to obtain a relation between $F$ and $F^+$ could rely on the self-consistency equations (\ref{matricesS}).
In order to avoid confusion,
we note that in general the pair potential can be directly calculated from $F$ and the interaction $V$, however, for the calculation of $F$ in terms of $\Delta$ in general a set of differential equations must be solved (the Gor'kov equations), which depends on additional interactions and, importantly, on boundary conditions.
If one decomposes the functions $F$ and $F^+$ into symmetry components, $F=\bar F+ \breve F$, where $\bar F$ has the same symmetry as $\Delta $, while $\breve F$ has differing symmetries [see, e.g., Eq.\ (\ref{F(k)})], then $\Delta_{\alpha\beta}=-V_{\alpha\beta;\gamma\delta}(1-2)\bar F_{\gamma\delta}(1,2)$ and $V_{\alpha\beta;\gamma\delta}(1-2)\breve F_{\gamma\delta}(1,2)=0$ hold.
Then, only for the component $\bar F$ can one obtain a relation
$\bar F^+_{\alpha\beta}(1,2) = \bar F_{\beta\alpha}^*(2,1)$,
finally leading to $\bar F^+_{\alpha\beta}(k) = \bar F_{\beta\alpha}^*(k)$. This relation
applies, e.g., to conventional bulk superconducting states, or the odd-$\omega$--dia state of Refs.~\cite{Solenov,Kusunose}.
However, it does not apply to the induced odd-$\omega$--para components of Refs.~\cite{BVE,Asano2007a,Nazarov07,Eschrig08}, since about the relation between $\breve F$ and $\breve F^+$ nothing can be inferred from the self-consistency equations.

We call the superconducting correlations with the same symmetry as the pair potential
{\it principal} components, while other components --- \textit{induced} (e.g., by additional interactions, interfaces, other inhomogeneities, or external fields). In other words, the principal components of $F$ and $F^+$ correspond to the symmetry components of $\Delta$, while induced components of $F$ and $F^+$ do not contribute to the right-hand sides of Eqs.\ (\ref{matricesS}) due to the structure of $V$. The odd-$\omega$--dia state discussed in Refs.~\cite{Solenov,Kusunose} was implicitly assumed to be the principal component.

\subsection{Homogeneous case}

In the homogeneous case, the self-consistency equation (\ref{matricesSa}) takes the form
\begin{equation}
\Delta_{\alpha\beta}(k) = - \int (dk') V_{\alpha\beta;\gamma\delta}(k-k') F_{\gamma\delta}(k').
\end{equation}
Relation (\ref{Delta12}) turns to (\ref{Delta+Delta*}) and is simplified as
\begin{equation} \label{Delta+Delta*hom}
\Delta^+_{\alpha\beta}(k) = \Delta^*_{\beta\alpha}(k).
\end{equation}

In the general case, the pair potential can be decomposed into the singlet component $d_0(k)$ and the triplet component $\mathbf d(k)$ as
\begin{equation} \label{Delta_components}
\Delta_{\alpha\beta}(k) = d_0(k) (i\sigma_2)_{\alpha\beta} + \mathbf d(k) (i \boldsymbol\sigma \sigma_2)_{\alpha\beta},
\end{equation}
where $\sigma_i$ with $i=1,2,3$ denotes the Pauli matrices in the spin space (below we will also use $\sigma_0$ to denote the unity matrix).
Fermionic antisymmetry of $\Delta$ [Eq.\ (\ref{Deltaantisym})] implies
\begin{align}
d_0(k) &= d_0(-k) = s_\Delta d_0(-\mathbf k,\omega), \\
\mathbf d(k) &= - \mathbf d(-k) = -s_\Delta \mathbf d(-\mathbf k,\omega).
\end{align}
When discussing principal odd-$\omega$ states with $s_\Delta=-1$ (recall that superconducting components of $\Delta$ are principal by definition), we {\it assume} that it can indeed be realized due to some proper interaction. The possibility to find such an interaction is not at all clear to date.

From Eq.\ (\ref{Minverse}) we then find the Green functions
%(assuming $\xi_{-\mathbf k} = \xi_{\mathbf k}$)
\cite{Kusunose}:
\begin{align}
G_{\alpha\beta}(k) &= - G'_{\beta\alpha}(-k) = G_0(k) \delta_{\alpha\beta} +\mathbf G(k) \boldsymbol\sigma_{\alpha\beta},
\label{GF1} \\
F_{\alpha\beta}(k) &= F^+_{\beta\alpha}(k)^* = F_0(k) (i\sigma_2)_{\alpha\beta} +\mathbf F(k) (i \boldsymbol\sigma \sigma_2 )_{\alpha\beta},
\label{GF2}
\end{align}
where the scalar and vector components are given by
\begin{align}
G_0(k) &= -\frac{(i\omega+\xi_{\mathbf k}) [ \omega^2+\xi_{\mathbf k}^2 +D_0(k)]}{\left[ \omega^2 +E_+^2(k) \right] \left[ \omega^2 +E_-^2(k) \right]},
\\
\mathbf G(k) &= \frac{(i\omega+\xi_{\mathbf k}) \mathbf D(k)}{\left[ \omega^2 +E_+^2(k) \right] \left[ \omega^2 +E_-^2(k) \right]},
\\
F_0(k) &= \frac{(\omega^2+\xi_{\mathbf k}^2) d_0(k) + [ d_0^2(k)-\mathbf d^2(k) ] d_0^*(k)}{\left[ \omega^2 +E_+^2(k) \right] \left[ \omega^2 +E_-^2(k) \right]},
\\
\mathbf F(k) &= \frac{(\omega^2+\xi_{\mathbf k}^2) \mathbf d(k) - [ d_0^2(k)-\mathbf d^2(k) ] \mathbf d^*(k)}{\left[ \omega^2 +E_+^2(k) \right] \left[ \omega^2 +E_-^2(k) \right]},
\end{align}
with
\begin{equation}
E_\pm(k) = \sqrt{\xi_{\mathbf k}^2 + D_0(k) \pm D(k)}.
\end{equation}
The real functions $D_0$, $\mathbf D$, and $D$ arise from the expression
\begin{equation}
\Delta(k) \Delta^+(k) = D_0(k) \hat 1 + \mathbf D(k) \boldsymbol\sigma,
\end{equation}
and have the following explicit forms:
\begin{align}
D_0(k) &= d_0^*(k) d_0(k) + \mathbf d^*(k) \mathbf d(k),
\label{D0Ddef} \\
\mathbf D(k) &= d_0(k) \mathbf d^* (k) + d_0^*(k) \mathbf d(k) + i [ \mathbf d(k) \times \mathbf d^*(k) ],
\notag \\
D(k) &= \sqrt{ \mathbf D^2(k)}. \notag
\end{align}

Note that relation $F^+_{\alpha\beta}(k) = F^*_{\beta\alpha}(k)$ [see Eq.\ (\ref{GF2})] is realized here since we explicitly consider the superconducting state originating from Eqs.\ (\ref{S0fermwithspin_s}) and (\ref{Sintspins}), and having only principal components of the anomalous averages.

\subsection{Meissner kernel and superfluid density}

Assuming the London gauge, $\mathbf q \mathbf A(q)=0$, we find the linear response of the current to the external vector potential in the form
\begin{equation} \label{jKA}
j_i(q) = -\frac 1c \mathcal K_{ij}(q) A_j(q),
\end{equation}
with the Meissner kernel \cite{Solenov,Kusunose}
\begin{multline} \label{Kkernel}
\mathcal K_{ij}(q)
= \frac{e^2}{m} \int (dk)
\frac{k_i k_j}m
\Bigl[
G_{\alpha\beta}(k) G_{\beta\alpha}(k-q)
\\
+
F_{\alpha\beta}(k) F^+_{\beta\alpha}(k-q)
\Bigr]
+
\frac{n e^2}m \delta_{ij},
\end{multline}
where $q=(\mathbf q,\epsilon_l)$ with the bosonic Matsubara frequency, $\epsilon_l = 2l\pi T$, and $n$ is the electronic density.

The tensor structure of the kernel depends on the orbital symmetry of the superconducting state. We discuss here for brevity isotropic (\textit{s}-wave) superconducting states, when the Green functions do not depend on the direction of the wave vector. Integrating over $d^3 \mathbf k$ in Eq.\ (\ref{Kkernel}), we choose $\mathbf q$ as the polar axis $z$ of the spherical coordinate system. Then, due to integration in the azimuthal plane all nondiagonal components of $\mathcal K_{ij}(q)$ vanish. The tensor is thus diagonal, with $\mathcal K_{xx}=\mathcal K_{yy}$ due to symmetry, while $\mathcal K_{zz}$ is generally different. At the same time, due to the London gauge, $\mathbf A(q)$ does not have a component along $\mathbf q$, so that the current (\ref{jKA}) is insensitive to the $\mathcal K_{zz}$ component, which allows us to choose it arbitrarily. For simplicity, we take
\begin{equation}
\mathcal K_{ij}(q) = \mathcal K(q) \delta_{ij}.
\end{equation}

The integral in Eq.\ (\ref{Kkernel}) is divergent, and we employ the usual trick regularizing the divergency by subtracting the normal-metal expression [since $\mathcal K(q)=0$ in the normal state]:
\begin{multline} \label{K(q)}
\mathcal K(q)
= \frac{\mathcal K_{ii}(q)}3
= \frac{e^2}{3m} \int (dk)
\frac{\mathbf k^2}m
\Bigl[
G_{\alpha\beta}(k) G_{\beta\alpha}(k-q)
\\
+
F_{\alpha\beta}(k) F^+_{\beta\alpha}(k-q)
-
G_{\alpha\beta}^{(0)}(k) G_{\beta\alpha}^{(0)}(k-q)
\Bigr],
\end{multline}
here the $(0)$ superscript denotes the normal-metallic functions.

At $q=0$, the kernel gives the superfluid density $n_S$:
\begin{equation} \label{K(0)}
\mathcal K(0) = \frac{e^2 n_S}m.
\end{equation}

Taking into account the explicit spin structure of the Green functions (\ref{GF1}) and (\ref{GF2}), we find
\begin{multline} \label{nS}
\frac{n_S}n = T \sum_\omega \int_{-\infty}^\infty d\xi
\Bigl[
G_0^2(\xi,\omega) + \mathbf G^2(\xi,\omega)
\\
+
F_0^*(\xi,\omega) F_0(\xi,\omega) + \mathbf F^*(\xi,\omega) \mathbf F(\xi,\omega)
\Bigr].
\end{multline}
Note the positive contribution from the anomalous components of the Green function, which stems from
relation $F^+_{\alpha\beta}(k) = F^*_{\beta\alpha}(k)$ [see Eq.\ (\ref{GF2})], which is realized here since we explicitly consider the superconducting state with only principal components of the anomalous averages.

\subsection{Principal \textit{s}-wave singlet even-\texorpdfstring{$\omega$}{w} state}
In the \textit{s}-wave singlet (hence, even-$\omega$) case, the triplet component $\mathbf d(k)$ is zero, and we have
\begin{equation}
G_0 = -\frac{i\omega +\xi}{\omega^2 +\xi^2 +|d_0|^2},
\quad
F_0 = \frac{d_0}{\omega^2 +\xi^2 +|d_0|^2},
\end{equation}
where $d_0 = d_0(|\mathbf k|,\omega)$ is an even function of $\omega$.
Even though $d_0$ depends in general on $|\mathbf k|$, we make the usual approximation of weak dependence
near the Fermi surface, so that during integration over $\xi$ in Eq.\ (\ref{nS}) we can fix $|\mathbf k|=k_F$.
Then, we reproduce the standard result
\begin{equation} \label{nSsinglet}
\frac{n_S}n = \pi T \sum_\omega \frac{|d_0|^2}{\left( \omega^2 +|d_0|^2 \right)^{3/2}},
\end{equation}
and the superfluid density is positive.

\subsection{Principal \textit{s}-wave triplet odd-\texorpdfstring{$\omega$}{w}--dia state}
In the \textit{s}-wave triplet (hence, odd-$\omega$) case, $d_0=0$ while $\mathbf d = \mathbf d(|\mathbf k|,\omega)$ is an odd function of $\omega$, and we have
\begin{align}
G_0 &= -\frac{(i\omega+\xi) ( \omega^2+\xi^2 +D_0)}{\left( \omega^2 +\xi^2 +D_0 +D \right) \left( \omega^2 +\xi^2 +D_0 -D \right)},
\label{GFoddwdia1} \\
\mathbf G &= \frac{(i\omega+\xi) \mathbf D}{\left( \omega^2 +\xi^2 +D_0 +D \right) \left( \omega^2 +\xi^2 +D_0 -D \right)},
\label{GFoddwdia2} \\
F_0 &= 0,
\label{GFoddwdia3} \\
\mathbf F &= \frac{\omega^2+\xi^2 +D_0}{\left( \omega^2 +\xi^2 +D_0 +D \right) \left( \omega^2 +\xi^2 +D_0 -D \right)} \mathbf d,
\label{GFoddwdia4}
\end{align}
where
\begin{equation} \label{D0D}
D_0 = \mathbf d^* \mathbf d,\quad \mathbf D = i\left[ \mathbf d \times \mathbf d^* \right],\quad D =\sqrt{D_0^2- (\mathbf{dd})^* (\mathbf{dd})}.
\end{equation}
We again approximate $\mathbf d(\mathbf k)$ by its value at the Fermi surface.
Then, Eq.\ (\ref{nS}) yields
\begin{multline} \label{nSoddwdia}
\frac{n_S}n = \pi T \sum_\omega \frac 1{8 D}
\biggl[
\frac{2D_0 \omega^2 + 2D_0^2 -2 D^2 + D_0 D}{\left( \omega^2+D_0 - D \right)^{3/2}}
\\
-
\frac{2D_0 \omega^2 + 2D_0^2 -2 D^2 - D_0 D}{\left( \omega^2+D_0 + D \right)^{3/2}}
\biggr].
\end{multline}
Taking into account that $D \leqslant D_0$ [see Eq.\ (\ref{D0D})], one can check that the expression in the square brackets is always non-negative,
so $n_S>0$.
In the case of unitary pairing, i.e., $\mathbf D=0$, the result has a form similar to the singlet case (\ref{nSsinglet}):
\begin{equation} \label{nSoddwdia_un}
\frac{n_S}n = \pi T \sum_\omega \frac{\mathbf d^* \mathbf d}{\left( \omega^2 +\mathbf d^* \mathbf d \right)^{3/2}}.
\end{equation}
Thus, the superfluid density is positive also for the \textit{s}-wave triplet odd-$\omega$ case.

\subsection{Induced \textit{s}-wave triplet odd-\texorpdfstring{$\omega$}{w}--para state}

The odd-$\omega$ state discussed above [Eqs.\ (\ref{GFoddwdia1})--(\ref{nSoddwdia_un})] demonstrates diamagnetic response to the external magnetic field ($n_S>0$), and, according to Refs.~\cite{Solenov,Kusunose}, is therefore consistent as a principal superconducting state (while the question of finding a proper interaction is to date unclear). Below, we discuss the induced \textit{s}-wave triplet odd-$\omega$--para state (which is undoubtedly realized, e.g., in S/F \cite{BVE,Asano2007a,Nazarov07,Eschrig08}
or TS/DN \cite{Tanaka,Higashitani09,Asano2013,Tanaka2004} proximity structures)
using the same language.

Since the odd-$\omega$--para state is induced, it does not have a pair potential in the corresponding symmetry channel, while the corresponding superconducting correlations are described by anomalous averages. In order to use the language established above, we can still use notations $\Delta$ and $\Delta^+$, however, consider them now simply as auxiliary quantities parametrizing the Green functions. The Green functions found in microscopic models are, of course, model-dependent, but here we are interested in fundamental properties determined by symmetries. For an example of a simple microscopic model, see Appendix~\ref{sec:app:sF}.

As has been previously discussed \cite{Solenov,Kusunose}, the odd-$\omega$--dia and --para states are characterized by different signs in the relation between $F$ and $F^+$. Instead of Eq.\ (\ref{FF+hom}) with $s_\Delta=-1$ (odd-$\omega$--dia state), one has $s_\Delta=1$ in this relation
for the odd-$\omega$--para state (this state is realized in microscopic models with even-$\omega$ pair potentials). In order to capture this property, we parametrize the Green functions taking the same form (\ref{Delta_components}) for $\Delta_{\alpha\beta}(k)$, and define $\Delta^+$ according to
\begin{equation} \label{Deltadefin}
\Delta^+_{\alpha\beta}(1,2) \equiv -\Delta_{\beta\alpha}^*(1,2),
\quad
\Delta^+_{\alpha\beta}(k) = -\Delta_{\beta\alpha}^*(k);
\end{equation}
note that the signs are different from Eqs.\ (\ref{Delta12}) and (\ref{Delta+Delta*hom}) (which were valid, in particular, for a principal odd-$\omega$--dia state). The Green functions can then be found from Eqs.\ (\ref{M}) and (\ref{Minverse}). The result can be obtained from Eqs.\ (\ref{GFoddwdia1})--(\ref{GFoddwdia4}) by inverting the signs in front of $D_0$ and $\mathbf D$ in all expressions.

Instead of Eq.\ (\ref{nSoddwdia}) we now have
\begin{multline} \label{nSoddwpara}
\frac{n_S}n = \pi T \sum_\omega \frac 1{8 D}
\biggl[
\frac{-2D_0 \omega^2 + 2D_0^2 -2 D^2 - D_0 D}{\left( \omega^2-D_0 - D \right)^{3/2}}
\\
-
\frac{-2D_0 \omega^2 + 2D_0^2 -2 D^2 + D_0 D}{\left( \omega^2-D_0 + D \right)^{3/2}}
\biggr].
\end{multline}
We assume that the $\omega$ dependence of $D_0$ and $D$ is such that $\omega >\sqrt{D_0- D}$, so that the combinations in the denominators of Eq.\ (\ref{nSoddwpara}) are positive (it can be verified that this phenomenological assumption corresponds to existing microscopic models for the odd-$\omega$--para state).

Taking into account that $D \leqslant D_0$ [see Eq.\ (\ref{D0D})], one obtains that the expression in the square brackets in Eq.\ (\ref{nSoddwpara}) is always non-positive,
so $n_S<0$ (paramagnetic response).
In the case of unitary pairing, i.e., $\mathbf D=0$, we find
\begin{equation}
\frac{n_S}n = -\pi T \sum_\omega \frac{\mathbf d^* \mathbf d}{\left( \omega^2 -\mathbf d^* \mathbf d \right)^{3/2}}.
\end{equation}

\section{Coexistence of odd-\texorpdfstring{$\omega$}{w}--dia and odd-\texorpdfstring{$\omega$}{w}--para states}
\label{sec:coexistence}

\subsection{Superfluid density}
\label{sec:bulk}

Now, we consider a possibility of coexistence of odd-$\omega$--dia and --para states in some region of space. We assume that the Green functions are linear combinations of the two contributions:
\begin{align}
G_{\alpha\beta} &= (G_{d})_{\alpha\beta} + (G_{p})_{\alpha\beta}, \\
F_{\alpha\beta} &= (F_{d})_{\alpha\beta} + (F_{p})_{\alpha\beta}, \\
F^+_{\alpha\beta} &= (F^+_{d})_{\alpha\beta} + (F^+_{p})_{\alpha\beta},
\end{align}
where the dia and para contributions, in accordance with our previous consideration, have the following properties:
\begin{align}
(F^+_{d})_{\alpha\beta}(k) &= (F_{d})_{\beta\alpha}^*(k), \label{F+d} \\
(F^+_{p})_{\alpha\beta}(k) &= -(F_{p})_{\beta\alpha}^*(k). \label{F+p}
\end{align}
Instead of Eq.\ (\ref{nS}), from Eqs.\ (\ref{K(q)}) and (\ref{K(0)}) we now find
\begin{multline}
\frac{n_S}n = T\sum_\omega \int_{-\infty}^\infty d\xi
\Bigl[
(G_{0d} + G_{0p})^2 + (\mathbf G_d + \mathbf G_p)^2
\\
+
( F_{0d} + F_{0p} )( F_{0d}^* - F_{0p}^* ) + ( \mathbf F_d + \mathbf F_p )( \mathbf F_d^* - \mathbf F_p^* )
\Bigr] .
\end{multline}
This result contains separate contributions from the dia and para states, as well as the cross term
\begin{multline} \label{cross}
\frac{\delta n_S}n = T\sum_\omega \int_{-\infty}^\infty d\xi
\bigl[
2 G_{0d} G_{0p} + 2 \mathbf G_d \mathbf G_p
\\
+
( F_{0d}^* F_{0p} - F_{0d} F_{0p}^* ) + ( \mathbf F_d^* \mathbf F_p - \mathbf F_d \mathbf F_p^* )
\bigr].
\end{multline}
We see that the contribution from the anomalous functions is purely imaginary, so the cross term is complex-valued.

Considering unitary pairing for simplicity, we have
\begin{gather}
G_{0d} = -\frac{i\omega +\xi}{\omega^2 +\xi^2 + D_{0d}},
\quad
G_{0p} = -\frac{i\omega +\xi}{\omega^2 +\xi^2 - D_{0p}},
\label{oddwdiapara_b} \\
\mathbf G_d = \mathbf G_p =0,
\quad
F_{0d} = F_{0p} =0, \\
\mathbf F_{d} = \frac{\mathbf d_d}{\omega^2 +\xi^2 + D_{0d}},
\quad
\mathbf F_{p} = \frac{\mathbf d_p}{\omega^2 +\xi^2 - D_{0p}},
\label{oddwdiapara_e}
\end{gather}
and then
\begin{multline} \label{cross1}
\frac{\delta n_S}n = \pi T \sum_\omega
\frac 1{D_{0d}+D_{0p}}
\Biggr[
\frac{4\omega^2 + 2D_{0d} + \mathbf d_d \mathbf d_p^* - \mathbf d_d^* \mathbf d_p}{\sqrt{\omega^2 + D_{0d}}}
\\
-
\frac{4\omega^2 - 2D_{0p} + \mathbf d_d \mathbf d_p^* - \mathbf d_d^* \mathbf d_p}{\sqrt{\omega^2 - D_{0p}}}
\Biggr].
\end{multline}
This expression is complex due to the purely imaginary combination $(\mathbf d_d \mathbf d_p^* - \mathbf d_d^* \mathbf d_p)$.

A complex-valued cross term would mean a \emph{complex current} and is therefore unphysical. We conclude that assuming a possible coexistence of odd-$\omega$--dia and odd-$\omega$--para states, we arrive at an unphysical result.

\subsection{Josephson junction}
\label{sec:Josephson}

Now, we consider a Josephson junction in the tunneling limit, assuming that the banks are represented by odd-$\omega$ superconducting states. Our main interest is the combination of dia and para states.

The tunneling contribution to the action has the standard form,
\begin{multline}
\mathcal{S}_T = \int d\tau d\mathbf{r}_L d\mathbf{r}_R
\bigl[ \mathcal{T}_{\mathbf{r}_L \mathbf{r}_R} \psi_{L\alpha}^* (\mathbf{r}_L,\tau) \psi_{R\alpha} (\mathbf{r}_R,\tau)
\\
+ \mathcal{T}_{\mathbf{r}_L \mathbf{r}_R}^* \psi_{R\alpha}^* (\mathbf{r}_R,\tau) \psi_{L\alpha}
(\mathbf{r}_L,\tau) \bigr],
\end{multline}
with the tunneling matrix element $\mathcal{T}_{\mathbf{r}_L \mathbf{r}_R}$.

In addition to the particle-hole (PH) and spin spaces, we now also have the left-right (LR) space, so we deal with the direct product of the three spaces:
PH $\otimes$ spin $\otimes$ LR.
In order to write the quadratic (over fermions) part of the action in a compact form, we define the vector field
\begin{equation}
\Psi_\alpha = \left(
\psi_{L\alpha} , \psi_{L\alpha}^* , \psi_{R\alpha} , \psi_{R\alpha}^*
\right)^\mathrm{T}.
\end{equation}
Then, instead of Eq.\ (\ref{Squad}) we can write the fermion part of the action as
\begin{gather}
\mathcal{S}_0 +\mathcal{S}_\mathrm{aux} + \mathcal{S}_T = \frac 12 \int_{12}
\Psi_\alpha^\dagger(1)
 \begin{pmatrix}
 \hat M_{L\alpha\beta} & \hat{\mathcal{T}}_{LR\alpha\beta}  \\
 \hat{\mathcal{T}}_{RL\alpha\beta} & \hat M_{R\alpha\beta}
 \end{pmatrix}
\Psi_\beta(2) ,
\\
\hat{\mathcal{T}}_{LR\alpha\beta} =
\begin{pmatrix}
\mathcal{T}_{\mathbf r_1 \mathbf r_2} & 0 \\
0 & -\mathcal{T}_{\mathbf r_1 \mathbf r_2}^*
\end{pmatrix}_\mathrm{PH}
\delta_{\alpha\beta}
\delta(\tau_1-\tau_2),
\\
\hat{\mathcal{T}}_{RL\alpha\beta} =
\begin{pmatrix}
\mathcal{T}_{\mathbf r_2 \mathbf r_1}^* & 0 \\
0 & -\mathcal{T}_{\mathbf r_2 \mathbf r_1}
\end{pmatrix}_\mathrm{PH}
\delta_{\alpha\beta}
\delta(\tau_1-\tau_2).
\end{gather}
After integrating over fermions we find
\begin{equation} \label{withboundary}
\mathcal{S}_0 +\mathcal{S}_\mathrm{aux} + \mathcal{S}_T = - \frac 12 \Tr \ln
 \begin{pmatrix}
  \check M_L  & \check{\mathcal{T}}_{LR} \\
  \check{\mathcal{T}}_{RL} & \check M_R
 \end{pmatrix},
\end{equation}
where the matrix under the logarithm is written explicitly in the LR space, while its elements are $4\times 4$ matrices in the PH $\otimes$ spin space.

In the tunneling limit we expand the logarithm over the nondiagonal (in the LR space) part. The
tunneling contribution arises in the second order:
\begin{equation} \label{S_T}
\mathcal{S}_T =
\frac 12 \Tr \left( \check M_L^{-1} \check{\mathcal{T}}_{LR} \check M_R^{-1} \check{\mathcal{T}}_{RL} \right).
\end{equation}
We make the standard assumption that the tunneling matrix element in the momentum representation $\mathcal{T}_{\mathbf k \mathbf k'}$ does not depend on the momenta (we denote this value by $\mathcal{T} =|\mathcal{T}| e^{i\alpha}$). Then,
\begin{multline} \label{STgen}
\mathcal{S}_T
= \frac 12 \sum_\omega \int (d^3 \mathbf k_L) (d^3 \mathbf k_R)
\\
\times \Tr \Bigl[
|\mathcal{T}|^2 \left\{ \hat G_{L} (\omega,\mathbf k_L) \hat G_{R} (\omega,\mathbf k_R)
+ \hat G'_{L} (\omega,\mathbf k_L) \hat G'_{R} (\omega,\mathbf k_R)\right\}
\\
- \mathcal{T}^{*2} \hat F_{L} (\omega,\mathbf k_L) \hat F^+_{R} (\omega,\mathbf k_R)
- \mathcal{T}^{2} \hat F^+_{L} (\omega,\mathbf k_L) \hat F_{R} (\omega,\mathbf k_R)
\Bigr],
\end{multline}
where the Green functions are written as matrices in the spin space.

For a junction between two conventional \textit{s}-wave singlet even-$\omega$ superconductors (with superconducting phases $\varphi_L$ and $\varphi_R$), we obtain:
\begin{equation}
\mathcal{S}_T
= -\frac{\pi G}2 \sum_\omega
\frac{\omega^2 + |d_{0L}| |d_{0R}| \cos\varphi}{\sqrt{\omega^2 +|d_{0L}|^2} \sqrt{\omega^2 +|d_{0R}|^2}},
\end{equation}
where $\varphi = (\varphi_R-\varphi_L +2\alpha)$ and $G$ is the interface conductance in units of $e^2/\hbar$,
\begin{equation}
G = 4\pi |\mathcal T|^2 \nu_L \nu_R,
\end{equation}
with $\nu_{L(R)}$ being the normal-metallic density of states in the L(R) superconductor.
The phase $\alpha$ of the tunneling matrix element only shifts the superconducting phase difference, so we can set $\alpha=0$ and then deal with the junction characterized by zero phase difference in equilibrium.
The anomalous part of the tunneling action describes Josephson coupling and can be written in terms of the critical current $I_c$:
\begin{equation}
\mathcal S_J = -\frac{I_c}{2e T} \cos\varphi,
\end{equation}
leading to the standard Josephson relation $I=I_c \sin\varphi$.

Considering a junction between two triplet unitary (for simplicity) odd-$\omega$ superconductors [see Eqs.\ (\ref{oddwdiapara_b})--(\ref{oddwdiapara_e})], we find
\begin{multline} \label{ST}
\mathcal{S}_T
= \sum_\omega \int (d^3 \mathbf k_L) (d^3 \mathbf k_R)
\\
\times
\Bigl[
2 |\mathcal T|^2 G_{0L} (\omega,\mathbf k_L) G_{0R} (\omega,\mathbf k_R)
\\
- s_R \mathcal T^{*2} \mathbf F_{L} (\omega,\mathbf k_L) \mathbf F^*_{R} (\omega,\mathbf k_R)
\\
- s_L \mathcal T^2 \mathbf F^*_{L} (\omega,\mathbf k_L) \mathbf F_{R} (\omega,\mathbf k_R)
\Bigr],
\end{multline}
where $s_L$ and $s_R$ are the signs originating from Eqs.\ (\ref{F+d}) and (\ref{F+p}) and defined as follows: $s_{L(R)} = \pm 1$ if the corresponding superconductor is odd-$\omega$-dia/para. Consequently, if both sides of the junction are of the same type, the Josephson (anomalous) part of the action is real, while if they are of different types, the Josephson contribution is purely imaginary.

The $\mathbf d$ vector for the unitary pairing can be represented as $\tilde{\mathbf d} e^{i\varphi}$ with a real vector $\tilde{\mathbf d}$.
Then, taking the Green functions given by Eqs.\ (\ref{oddwdiapara_b})--(\ref{oddwdiapara_e}), we obtain
\begin{equation} \label{dvect}
\mathcal{S}_T
= -\frac{\pi G}2 \sum_\omega
\frac{\omega^2 + \tilde{\mathbf d}_L \tilde{\mathbf d}_R (s_L e^{i\varphi} + s_R e^{-i\varphi})/2}{\sqrt{\omega^2 + s_L \tilde{\mathbf d}_L^2} \sqrt{\omega^2 +s_R \tilde{\mathbf d}_R^2}}.
\end{equation}
The phase-dependent combination in the anomalous part takes a form depending on the types of superconductors composing the junction:
\begin{equation} \label{sLsR}
\frac{s_L e^{i\varphi} + s_R e^{-i\varphi}}2
=
\left\{
\begin{array}{cl}
-\cos\varphi, & \text{odd-$\omega$--para/para,} \\
\cos\varphi, & \text{odd-$\omega$--dia/dia,} \\
-i\sin\varphi, & \text{odd-$\omega$--para/dia.}
\end{array}
\right.
\end{equation}
We can obtain equivalent results in the quasiclassical technique ($\xi$-integrated Green functions) with
proper boundary conditions \cite{Nazarov,KL}.

Strikingly, the last case (odd-$\omega$--para/odd-$\omega$--dia junction) yields {\it purely imaginary Josephson coupling}, leading to the Josephson current proportional to $i\cos\varphi$. This result is unphysical and signifies problems regarding possible coexistence of odd-frequency pairings with different types of Meissner response.

Summarizing the results of the present subsection, we see that while a conventional even-$\omega$/even-$\omega$ junction demonstrates the standard Josephson current proportional to $\sin\varphi$, the situation with odd-$\omega$ superconductors is more complicated. In even-$\omega$/odd-$\omega$ junctions, the first-order Josephson coupling is absent due to different spin structure of the banks. Same-type combinations odd-$\omega$--dia/odd-$\omega$--dia and odd-$\omega$--para/odd-$\omega$--para lead to sinusoidal current-phase relation (the general sign can be negative, which corresponds to the $\pi$ junction). At the same time, considering a different-type odd-$\omega$--para/odd-$\omega$--dia junction, we arrive at the Josephson current proportional to $i\cos\varphi$, which is imaginary and thus unphysical.
Similarly to Sec.~\ref{sec:bulk}, we conclude that the assumption of coexistence of an odd-$\omega$--dia and an odd-$\omega$--para state leads to unphysical results.

One could try to avoid the contradiction, assuming that in odd-$\omega$--para/dia junctions only configurations with $\tilde{\mathbf d}_L \perp \tilde{\mathbf d}_R$ are realized, so that the imaginary contribution vanishes in the first-order Josephson coupling in Eq.\ (\ref{dvect}). However, an argument of this sort (pair potentials adjusting to each other in a junction) could work if other configurations were energetically unfavorable (i.e., they would yield a free energy not in the minimum). In our case, they yield a complex free energy, hence are simply unphysical. Therefore, this argument can hardly solve the issue.

Calculations for a specific microscopic model of the odd-$\omega$--para state in Appendix~\ref{sec:app:sF} confirm the results of the present subsection.

\section{Discussion}
\label{sec:discussion}

Recently, it was demonstrated that in uniform systems with broken time-reversal symmetry \cite{Kusunose12} and in nonuniform systems with preserved time-reversal symmetry \cite{Asano2014}, even-$\omega$ and odd-$\omega$ superconducting components can mix with each other. At the same time, it turns out that the states are divided into two separate classes \cite{Kusunose12,Asano2014}: while the even-$\omega$--dia state can mix only with the odd-$\omega$--para state, the even-$\omega$--para state can mix only with the odd-$\omega$--dia state. The two classes seem to avoid coexistence between themselves.

From this point of view, when studying odd-$\omega$--para/odd-$\omega$--dia mixtures, we assume coexistence between representatives of the two different classes. Our results demonstrate that this leads to unphysical superconducting transport properties of the systems.

How can this contradiction be resolved?
Superconductivity belonging to the first class is realized, e.g., in conventional \textit{s}-wave superconductors (even-$\omega$--dia) and in proximity systems with conventional superconductors (even-$\omega$--dia and odd-$\omega$--para states), and these states are described by well-established microscopic models.
At the same time, the question of possible realization of superconductivity belonging to the second class is still open.
The contradiction arising from the assumption of coexistence between the two classes then raises doubts about the actual possibility to realize the second-class superconductivity (in particular, the odd-$\omega$--dia state).

On the other hand, if the odd-$\omega$--dia state cannot be realized, then what is possibly wrong in the arguments of Refs.~\cite{Solenov,Kusunose}, where existence of this state was proposed from the general viewpoint of symmetry and stability (without presenting explicit microscopic interaction leading to this state)? The authors of Refs.~\cite{Solenov,Kusunose} argue that the odd-$\omega$--dia state can be realized only in a system with strongly retarded interaction, and cannot be described by a mean-field Hamiltonian. Assuming existence of a mean-field Hamiltonian immediately leads to Eq.\ (\ref{FF+hom}) with $s_\Delta=1$ that signifies the odd-$\omega$--para state \cite{Solenov,Kusunose}.
On the other hand, an effective retarded low-energy theory
in the path-integral formulation [Eqs.\ (\ref{Z})--(\ref{Sintspinsgen}) or (\ref{Z_s})--(\ref{Sintspins})] emerges after integrating out high-energy degrees of freedom in a many-body Hamiltonian. Assuming some general initial many-body Hamiltonian $\hat H$ (that, e.g., contains all electrons and nuclei of the solid with their mutual interactions),
one can define the Heisenberg operators
$\hat\psi_\alpha(\mathbf r,\tau) = e^{\hat H \tau} \hat\psi_\alpha(\mathbf r) e^{-\hat H \tau}$ and
$\hat\psi_\alpha^+ (\mathbf r,\tau) = e^{\hat H \tau} \hat\psi_\alpha^\dagger (\mathbf r) e^{-\hat H \tau}$,
and the anomalous Green functions
\begin{subequations} \label{FFmb}
\begin{align}
F_{\alpha\beta}(\mathbf r_1,\tau_1;\mathbf r_2,\tau_2) &= \bigl< T_\tau \hat\psi_\alpha(\mathbf r_1,\tau_1) \hat\psi_\beta(\mathbf r_2,\tau_2) \bigr>,
\\
F^+_{\alpha\beta}(\mathbf r_1,\tau_1;\mathbf r_2,\tau_2) &= \bigl< T_\tau \hat\psi_\alpha^+ (\mathbf r_1,\tau_1) \hat\psi_\beta^+ (\mathbf r_2,\tau_2) \bigr>,
\end{align}
\end{subequations}
where averaging is over the exact many-body state.
We are interested in the general relation between $F$ and $F^+$, which can be obtained directly from definitions (\ref{FFmb}), following the same logic as discussed for the mean-field case in Refs.~\cite{Solenov,Kusunose}.
Applying this to
the homogeneous case, we immediately find the relation $F_{\alpha\beta}^+(\mathbf k,\omega)=F_{\beta\alpha}^*(\mathbf k,-\omega)$ between the many-body anomalous averages, which has exactly the same form as Eq.\ (\ref{FF+hom}) with $s_\Delta=1$.
This seems to be an unavoidable fundamental relation.
Therefore, in order to obtain the odd-$\omega$--dia state described by relation (\ref{FF+hom}) with $s_\Delta=-1$, one has to assume the impossibility of a Hamiltonian description of the system at \emph{any} level (both mean-field and many-body). This seems unnatural to us and increases our skepticism about a possible realization of the odd-$\omega$--dia state.

Having established a fundamental incompatibility between a Hamiltonian description of the system at any level and the  possibility of the odd-$\omega$--dia state, we still have to explain how relation (\ref{FF+hom}) with $s_\Delta=-1$ can be avoided if one starts from the path-integral formulation of Eqs.\ (\ref{Z_s})--(\ref{Sintspins}). We suppose that the key issue here is the spontaneous symmetry breaking. Taking a trial path for the pair potential instead of integration over the $\Delta^*$ and $\Delta$ fields [introduced for the Hubbard-Stratonovich transformation in Eq.\ (\ref{HS})], we arrive at the mean-field definitions of the anomalous averages, Eqs.\ (\ref{FbarF}). This step in the derivation assumes that while the free energy has a superconducting manifold of equivalent minima (with arbitrary superconducting phase), due to an infinitesimally small perturbation (which is not even explicitly considered) the system chooses some definite phase (thus breaking the gauge symmetry). This choice of a single point from the manifold corresponds to taking the mean-field value of $\Delta$ (the trial path) instead of integrating over this field.

At the same time, the symmetry-breaking (phase-fixing) perturbation is an essential issue. In the case of conventional superconductivity, we can write it explicitly in the second-quantized representation as $\int d\mathbf r \bigl[ \delta_0 \hat{\psi}_\uparrow^\dagger (\mathbf r) \hat{\psi}_\downarrow^\dagger(\mathbf r) + \delta_0^* \hat{\psi}_\downarrow(\mathbf r) \hat{\psi}_\uparrow(\mathbf r) \bigr]$, with $|\delta_0| \to 0$. This Hermitian term breaks the global gauge symmetry, setting the preferential value of the phase to be equal to the phase of $\delta_0$. On the other hand, for the odd-$\omega$--dia solution of Refs.~\cite{Solenov,Kusunose} [Eq.\ (\ref{FF+hom}) with $s_\Delta=-1$], our hypothesis is that there is no spontaneous physical perturbation (fluctuation) that can lead to the spontaneous gauge-symmetry breaking, thus fixing the phase of this state (this point of view correlates with the absence of a Hamiltonian description of the odd-$\omega$--dia state). Then, one has to retain integration over the superconducting phase, and the anomalous averages (\ref{FbarF}) vanish, so that Eq.\ (\ref{FF+hom}) with $s_\Delta=-1$ is trivially satisfied. In other words, the corresponding odd-$\omega$--dia minimum manifold can exist, but the symmetry cannot be spontaneously broken since there exists no physical fluctuation that could fix the phase.
Note that in a related problem of an odd-$\omega$--dia state in Ref.~\cite{BK}, the authors admit that the properties of the anomalous averages ``depend on the properties of the unphysical, time-dependent, symmetry-breaking field.''

\section{Conclusions}
\label{sec:conclusions}

We have considered physical consequences of the odd-$\omega$--dia superconducting state proposed recently in Refs.~\cite{Solenov,Kusunose}. This is the odd-frequency state with diamagnetic Meissner response, which has no stability issues in the bulk and could therefore be realized as a principal superconducting state (at the same time, a convincing demonstration of such a state in a microscopic model is still lacking). Assuming the possibility to realize this state (due to a proper retarded interaction), we have studied its coexistence with the odd-$\omega$--para superconducting state, which is known to be generated as an induced superconducting component in, e.g., S/F or TS/DN proximity structures. Calculating the superfluid density of the mixed odd-$\omega$--para/dia state and the Josephson current in the odd-$\omega$--para/dia junction, we find that the currents in both cases have imaginary contributions and are therefore unphysical.

Taking into account rigorous microscopic derivations of the odd-$\omega$--para state in a number of models, we thus encounter the question of an actual realizability of the odd-$\omega$--dia state. Further analysis shows that a realization of this state implies the absence of a Hamiltonian description
for the system at any (mean-field or many-body) level. Technically, the essential difference between the two states is described by different signs in the relation between the two anomalous averages, see Eq.\ (\ref{FF+hom}) that has $s_\Delta=1$ for the odd-$\omega$--para and $s_\Delta=-1$ for the odd-$\omega$--dia state. We conclude that in the latter case there is no physical perturbation leading to the global phase-symmetry breaking, and in the absence of a physical perturbation in a mean-field Lagrangian formalism, Eq.\ (\ref{FF+hom}) with $s_\Delta=-1$ is trivially satisfied because the anomalous averages vanish due to integration over the superconducting phase.
This also seems to be the only way to avoid unphysical imaginary current
components in such a hypothetical odd-$\omega$--dia superconductor when brought in contact with the well-established induced odd-$\omega$--para state in, e.g.,  S/F or TS/DN hybrid structures.

\acknowledgments

We are grateful to A.~A.\ Golubov, P.~A.\ Ioselevich, M.~S.\ Kalenkov, H.\ Kusunose, and D.\ Mozyrsky for helpful discussions.
This work was partially supported by the ``Topological Quantum Phenomena'' (Nos.\ 22103002 and 22103005)
Grant-in-Aid for Scientific Research on Innovative Areas and
KAKENHI (No.\ 26287069) from the Ministry of Education,
Culture, Sports, Science and Technology (MEXT) of Japan,
by the RFBR-JSPS Grant No.\ 15-52-50054,
and by the Ministry of Education and Science of the Russian Federation (Grant No.\ 14Y.26.31.0007).
Ya.V.F.\ was supported in part by the program ``Quantum mesoscopic and disordered structures'' of the RAS.
M.E.\ acknowledges support from the EPSRC under Grant No.\ EP/J010618/1, as well as inspiring discussions at the Aspen Center of Physics and within the Hubbard Theory Consortium.

\appendix

\section{Fourier transformation}
\label{sec:app:Fourier}

The Fourier transformation is introduced according to the following rules:
\begin{align}
A(1,2) &= \int (dk) \int (dk') A(k,k') e^{i(k x_1 - k' x_2)},
\\
A(k,k') &= \int_1 \int_2 A(1,2) e^{-i(k x_1 - k' x_2)},
\end{align}
where
\begin{gather}
k=(\mathbf k,\omega),\quad kx \equiv \mathbf{kr}-\omega \tau,
\\
\int (dk)(\ldots) \equiv T \sum_\omega \int \frac{d\mathbf k}{(2\pi)^3}(\ldots).
\end{gather}
The fermionic Matsubara frequencies are $\omega= \pi T(2n+1)$.
According to these definitions,
\begin{equation}
\int_1 e^{-i (k-k') x_1} = \frac{\delta_{nn'}}T (2\pi)^3 \delta(\mathbf k -\mathbf k') \equiv \delta\left( \frac{k-k'}{2\pi} \right).
\end{equation}

So, for the Fourier transforms of the Green functions and the pair potentials we have
\begin{align}
G_{\alpha\beta}(k,k') &= -G'_{\beta\alpha}(-k',-k)
\notag\\
&= -\bigl< \psi_\alpha(k) \psi_\beta^*(k') \bigr>_\mathrm{MF}, \label{Gkk} \\
F_{\alpha\beta}(k,k') &= \bigl< \psi_\alpha(k) \psi_\beta(-k') \bigr>_\mathrm{MF}, \\
F^+_{\alpha\beta}(k,k') &= \bigl< \psi_\alpha^*(-k) \psi_\beta^*(k') \bigr>_\mathrm{MF}, \\
\Delta^+_{\alpha\beta}(k,k') &= \Delta^*_{\beta\alpha}(k',k). \label{Delta+Delta*}
\end{align}

\section{``Superconducting ferromagnet'' as a microscopic model of the odd-\texorpdfstring{$\omega$}{w}--para state}
\label{sec:app:sF}

In Sec.~\ref{sec:Josephson}, we discuss Josephson junctions between odd-$\omega$--para and odd-$\omega$--dia states introducing the former in a phenomenological manner with the help of auxiliary quantities $\Delta$ and $\Delta^+$, in order to use the same language as for the odd-$\omega$--dia case. We require Eqs.\ (\ref{Deltadefin}) to be fulfilled, so that the resulting state reproduces the symmetry and properties of the odd-$\omega$--para state known from microscopic models. At the same time, since clear microscopic models of the odd-$\omega$--para state are available, in this Appendix we check our general conclusions taking a specific microscopic model.

The simplest microscopic example of the odd-$\omega$--para state is realized in a ``superconducting ferromagnet,'' which is the conventional singlet superconductor (pair potential $\Delta$) with homogeneous exchange field $h$. The exchange field leads to appearance of the (induced) \textit{s}-wave triplet odd-$\omega$--para component, in addition to the conventional \textit{s}-wave singlet even-$\omega$--dia one.

Instead of Eq.\ (\ref{S0fermwithspin_s}) we now have
\begin{equation}
\mathcal S_0 = \int \left[ \psi_\alpha^* \left( \partial_\tau + \xi \right ) \psi_\alpha + h( \psi_\uparrow^* \psi_\uparrow - \psi_\downarrow^* \psi_\downarrow ) \right],
\end{equation}
and instead of Eq.\ (\ref{M}) we obtain
\begin{widetext}
\begin{equation}
\hat M_{\alpha\beta} (1,2)
=
\begin{pmatrix}
\delta(1-2) \left[ \delta_{\alpha\beta} (\partial_\tau+\xi) + h (\sigma_3)_{\alpha\beta} \right] &
i\Delta (\sigma_2)_{\alpha\beta}
\\
-i\Delta^* (\sigma_2)_{\alpha\beta} &
\delta(1-2) \left[ \delta_{\alpha\beta} (\partial_\tau-\xi) - h (\sigma_3)_{\alpha\beta} \right]
\end{pmatrix}.
\end{equation}
The Green functions (matrices in the spin space) are then given by
\begin{gather}
\hat G(k) = -\frac{\left[ (i\omega +\xi)(\omega^2 +\xi^2 +|\Delta|^2 - h^2) + 2i\omega h^2 \right] \hat\sigma_0
- h \left[ \omega^2+\xi^2+|\Delta|^2 -h^2 +2i\omega (i\omega +\xi) \right] \hat\sigma_3}
{(\omega^2 +\xi^2 +|\Delta|^2 - h^2)^2 + (2\omega h)^2},
\\
\hat F(k) = \hat F_s(k) + \hat F_t(k),
\qquad
\hat{F}^+(k) = \hat F_s^\dagger(k) - \hat F_t^\dagger(k),
\label{F(k)}
\end{gather}
\end{widetext}
where the singlet and triplet parts of the anomalous Green function have the following form:
\begin{align}
\hat F_s(k) &= \frac{(\omega^2 +\xi^2 +|\Delta|^2 - h^2)\Delta}
{(\omega^2 +\xi^2 +|\Delta|^2 - h^2)^2 + (2\omega h)^2}
(i\hat\sigma_2),
\\
\hat F_t(k) &= \frac{-2i\omega h \Delta}
{(\omega^2 +\xi^2 +|\Delta|^2 - h^2)^2 + (2\omega h)^2}
(i \hat\sigma_3 \hat\sigma_2).
\label{F_t}
\end{align}
Note that $\hat F_s(k)$ and $\hat F_t(k)$ have even-$\omega$/odd-$\omega$ symmetry, respectively, and Eqs.\ (\ref{F(k)}) are consistent with the general relation (\ref{FF+hom}) at $s_\Delta=1$ (since the pair potential belongs to the even-$\omega$ symmetry class).

Now, we consider a Josephson junction taking the above superconducting ferromagnet as the left bank of our junction, with $\Delta = |\Delta| e^{i\varphi_L}$. At the same time, we assume that the right bank of the junction is the principal \textit{s}-wave triplet odd-$\omega$--dia state, determined by Eqs.\ (\ref{GFoddwdia1})--(\ref{GFoddwdia4}). For simplicity, we assume the unitary case with $\mathbf d =\tilde{\mathbf d} e^{i\varphi_R}$ and a real vector $\tilde{\mathbf d}$, which is an odd function of $\omega$. Then,
\begin{gather}
G_0 = -\frac{i\omega +\xi}{\omega^2+\xi^2+ \tilde{\mathbf d}^2},
\quad
\mathbf F = \frac{\mathbf d}{\omega^2+\xi^2+ \tilde{\mathbf d}^2},
\\
\hat F = \mathbf F (i\hat{\boldsymbol\sigma} \hat\sigma_2),\quad \hat F^+ = \mathbf F^* (i\hat{\boldsymbol\sigma} \hat\sigma_2)^\dagger.
\end{gather}

Although the anomalous Green functions of the left bank, Eqs.\ (\ref{F(k)}), are mixtures of the even-$\omega$ and odd-$\omega$ components, the lowest-order Josephson coupling in the junction is provided only by the triplet component from the left bank, so that the Josephson coupling is effectively between the odd-$\omega$--para and --dia states.

The Josephson action [the anomalous part of the tunneling action (\ref{STgen})]
takes the form
\begin{multline} \label{SJferroS}
\mathcal S_J = -i \frac{\pi G}2 \cos\varphi
\\
\times \sum_\omega \Im \left( \frac{|\Delta|}{\sqrt{(\omega+ih)^2 + |\Delta|^2}} \right)
\frac{\tilde d_3(\omega)}{\sqrt{\omega^2+\tilde d_3^2(\omega)}}.
\end{multline}
Both the factors under the sum in the right-hand side are odd functions of $\omega$, so that their product is even and the sum is nonzero.
Note that the phase dependence in the case of this odd-$\omega$--para/dia junctions is determined by $\cos\varphi$ [in contrast to the last line in Eq.\ (\ref{sLsR})], because the phase of the odd-$\omega$--para component in the superconducting ferromagnet is shifted by $\pi/2$ with respect to the phase of $\Delta$ [note additional $i$ in the numerator in Eq.\ (\ref{F_t})].

At the same time, the main qualitative result of Eq.\ (\ref{SJferroS}) is that the general consideration of Sec.~\ref{sec:Josephson} is confirmed and the Josephson action for the odd-$\omega$--para/dia junction turns out to be purely imaginary and thus unphysical.


\begin{thebibliography}{99}

\bibitem{Mineev}
V.~P.\ Mineev and K.~V.\ Samokhin, \textit{Introduction to Unconventional Superconductivity} (Gordon and Breach,
London, 1999).

\bibitem{Berez}
V.~L.\ Berezinskii, Pis'ma Zh.\ Eksp.\ Teor.\ Fiz.\ \textbf{20}, 628 (1974) [JETP Lett.\ \textbf{20}, 287 (1974)].

\bibitem{KB}
T.~R.\ Kirkpatrick and D.\ Belitz, Phys. Rev. Lett. \textbf{66}, 1533 (1991); D.\ Belitz and T.~R.\ Kirkpatrick, Phys.\ Rev.~B \textbf{46}, 8393 (1992).

%%%%%%%%%%%%%%%%%%%%%%%%%%%%%%%%%%%%%%%%%%%%%%%%%%%%%%%%%%%%%
% Pairing mechanisms up to now
%
%%%%%%%%%%%%%%%%%%%%%%%%%%%%%%%%%%%%%%%%%%%%%%%%%%%%%%%%%%%%%%%%

%%%%%%%%%%%%%%%%%%%%%%%%%%%%%%%%%%%%%%%%%%%%%%%%%%%%%%%%%%%%%%%%
% Gapless pairing
%%%%%%%%%%%%%%%%%%%%%%%%%%%%%%%%%%%%%%%%%%%%%%%%%%%%%%%%%%%%%%%%%
\bibitem{Balatsky92}
A.\ Balatsky and E.\ Abrahams, Phys.\ Rev.~B \textbf{45}, 13125 (1992).

%%%%%%%%%%%%%%%%%%%%%%%%%%%%%%%%%%%%%%%%%%%%%%%%%%%%%%%%%%
% Composite Parings
%%%%%%%%%%%%%%%%%%%%%%%%%%%%%%%%%%%%%%%%%%%%%%%%%%%%%%%%%%
\bibitem{Emery}
V.~J.\ Emery and S.\ Kivelson, Phys.\ Rev.~B \textbf{46}, 10812 (1992);
Phys.\ Rev.\ Lett.\ \textbf{71}, 3701 (1993).

%%%%%%%%%%%%%%%%%%%%%%%%%%%%%%%%%%%%%%%%%%%%%%%%%
% Composite pairings
%%%%%%%%%%%%%%%%%%%%%%%%%%%%%%%%%%%%%%%%%%%%%%%%%%%%%%%%%
\bibitem{Balatsky93}
A.~V.\ Balatsky and J.\ Bonca, Phys.\ Rev.~B \textbf{48}, 7445 (1993).

\bibitem{Abrahams}
E.\ Abrahams, A.\ Balatsky, J.~R.\ Schrieffer, and P.~B.\ Allen,
Phys.\ Rev.~B \textbf{47}, 513 (1993).

%%%%%%%%%%%%%%%%%%%%%%%%%%%%%%%%%%%%%%%%%%%%%
% Gapless pairing Inhomogeneous state
%%%%%%%%%%%%%%%%%%%%%%%%%%%%%%%%%%%%%%%%%%%%%%%
\bibitem{Coleman}
P.\ Coleman, E.\ Miranda, and A.\ Tsvelik, Phys.\ Rev.\ Lett.\ \textbf{70}, 2960 (1993);
Phys.\ Rev.~B \textbf{49}, 8955 (1994);
P.\ Coleman, A.\ Georges, and A.~M.\ Tsvelik, J.~Phys.: Condens.\ Matter \textbf{9}, 345 (1997).

%%%%%%%%%%%%%%%%%%%%%%%%%%%%%%%%%%%%%%%%%%
% Inhomogeneous state
%%%%%%%%%%%%%%%%%%%%%%%%%%%%%%%%%%%%%%%%%%%%%
\bibitem{Heid}
R.\ Heid, Z.~Phys.~B \textbf{99}, 15 (1995).

\bibitem{Abrahams95}
E.\ Abrahams, A.\ Balatsky, D.~J.\ Scalapino, and J.~R.\ Schrieffer, Phys.\ Rev.~B \textbf{52}, 1271 (1995).

\bibitem{Zacher}
O.\ Zachar, S.~A.\ Kivelson, and V.~J.\ Emery, Phys.\ Rev.\ Lett.\ \textbf{77}, 1342 (1996).

\bibitem{Jarrell}
M.\ Jarrell, H.\ Pang, and D.~L.\ Cox, Phys.\ Rev.\ Lett.\ \textbf{78}, 1996 (1997).

\bibitem{Martisovots}
V.\ Martisovits and D.~L.\ Cox, Phys.\ Rev.~B \textbf{57}, 7466 (1998);
V.\ Martisovits, G.\ Zarand, and D.~L.\ Cox, Phys.\ Rev.\ Lett.\ \textbf{84}, 5872 (2000).

\bibitem{Vojta}
M.\ Vojta and E.\ Dagotto, Phys.\ Rev.~B \textbf{59}, R713(R) (1999).

%
\bibitem{BK}
D.\ Belitz and T.~R.\ Kirkpatrick,
%\textit{Properties of spin-triplet, even-parity superconductors},
Phys.\ Rev.~B \textbf{60}, 3485 (1999).
%

\bibitem{Coleman99}
P.\ Coleman, A.~M.\ Tsvelik, N.\ Andrei, and H.~Y.\ Kee, Phys.\ Rev.~B \textbf{60}, 3608 (1999).

\bibitem{Anders}
F.~B.\ Anders, Phys.\ Rev.~B \textbf{66}, 020504(R) (2002);
Eur.\ Phys.\ J.~B \textbf{28}, 9 (2002).

\bibitem{Fuseya2003}
Y.\ Fuseya, H.\ Kohno, and K.\ Miyake, J.~Phys.\ Soc.\ Jpn.\ \textbf{72}, 2914 (2003).

\bibitem{Sakai2004}
S.\ Sakai, R.\ Arita, and H.\ Aoki, Phys.\ Rev.~B \textbf{70}, 172504 (2004).

\bibitem{Hotta}
T.\ Hotta, J.~Phys.\ Soc.\ Jpn.\ \textbf{78}, 123710 (2009).

\bibitem{Shigeta1d}
K.\ Shigeta, S.\ Onari, K.\ Yada, and Y.\ Tanaka, Phys.\ Rev.~B \textbf{79}, 174507 (2009);
K.\ Shigeta, Y.\ Tanaka, K.\ Kuroki, S.\ Onari, and H.\ Aizawa, Phys.\ Rev.~B \textbf{83}, 140509(R) (2011);
K.\ Shigeta, S.\ Onari, and Y.\ Tanaka, J.~Phys.\ Soc.\ Jpn.\ \textbf{82}, 104702 (2013).

\bibitem{KFM}
H.\ Kusunose, Y.\ Fuseya, and K.\ Miyake, J.~Phys.\ Soc.\ Jpn.\ \textbf{80}, 044711 (2011).

\bibitem{ShigetaAF}
K.\ Shigeta, S.\ Onari, and Y.\ Tanaka, Phys.\ Rev.~B \textbf{85}, 224509 (2012).


%%%%%%%%%%%%%%%%%%%%%%%%%%%%%%%%%%%
% Composite
%%%%%%%%%%%%%%%%%%%%%%%%%%%%%%%%%%%%%%%%%
\bibitem{Dahal}
H.~P.\ Dahal, E.\ Abrahams, D.\ Mozyrsky, Y.\ Tanaka, and A.~V.\ Balatsky, New J.~Phys. \textbf{11}, 065005 (2009).

\bibitem{Yanagi12}
Y.\ Yanagi, Y.\ Yamashita, and K.\ Ueda, J.~Phys.\ Soc.\ Jpn.\ \textbf{81}, 123701 (2012).


%%%%%%%%%%%%%%%%%%%%%%%%%%%%%%%%%%%%%%%%%%%%%%%%
% Composite 2-channel Kondo
%%%%%%%%%%%%%%%%%%%%%%%%%%%%%%%%%%%%%%%%%%
\bibitem{Hoshino}
S.\ Hoshino, J.\ Otsuki, and Y.\ Kuramoto, Phys.\ Rev.\ Lett.\ \textbf{107}, 247202 (2011);
S.\ Hoshino and Y.\ Kuramoto, \textit{ibid.} \textbf{112}, 167204 (2014);
S.\ Hoshino, Phys.\ Rev.~B \textbf{90}, 115154 (2014).

%%%%%%%%%%%%%%%%%%%%%%%%%%%%%%%%%%%%%%%%%%%%
% F/S junction
%%%%%%%%%%%%%%%%%%%%%%%%%%%%%%%%%%%%%%%%%%%%%%%%
\bibitem{BVE}
F.~S.\ Bergeret, A.~F.\ Volkov, and K.~B.\ Efetov,
Phys.\ Rev.\ Lett.\ \textbf{86}, 4096 (2001);
%{\it Long-Range Proximity Effects in Superconductor-Ferromagnet Structures}
Rev.\ Mod.\ Phys.\ \textbf{77}, 1321 (2005).

%%%%%%%%%%%%%%%%%%%%%%%%%%%%%%%%%%%%%%%%%%%%%%%%%
% Induced by the broken spin-rotational symmetry
%%%%%%%%%%%%%%%%%%%%%%%%%%%%%%%%%%%%%%%%%%%%%%%%%%%
\bibitem{Tokuyasu1988}
T.\ Tokuyasu, J.~A.\ Sauls, and D.\ Rainer,
Phys.\ Rev.~B \textbf{38}, 8823 (1988).
%{\it Proximity effect of a ferromagnetic insulator in contact with a superconductor}

\bibitem{Kadigrobov01}
A.\ Kadigrobov, R.~I.\ Shekhter, and M.\ Jonson,
Europhys.\ Lett.\ \textbf{54}, 394 (2001).
%{\it Quantum spin fluctuations as a source of long-range proximity effects in diffusive ferromagnet-superconductor structures }

\bibitem{Eschrig03}
M.\ Eschrig, J.\ Kopu, J.~C.\ Cuevas, and G.\ Sch\"on,
Phys.\ Rev.\ Lett.\ \textbf{90}, 137003 (2003).
%{\it Theory of Half-Metal/Superconductor Heterostructures}

\bibitem{Lofwander05}
T.\ L\"ofwander, T.\ Champel, J.\ Durst, and M.\ Eschrig,
Phys.\ Rev.\ Lett.\ \textbf{95}, 187003 (2005).
%{\it Interplay of Magnetic and Superconducting Proximity Effects in Ferromagnet-Superconductor-Ferromagnet Trilayers}

\bibitem{Eschrig07}
M.\ Eschrig, T.\ L\"{o}fwander, T.\ Champel, J.~C.\ Cuevas, J. Kopu, and G.\ Sch\"{o}n, J.~Low.\ Temp.\ Phys.\ \textbf{147}, 457 (2007).
%{\it Symmetries of Pairing Correlations in Superconductor-Ferromagnet Nanostructures}

\bibitem{Nazarov07}
V.\ Braude and Yu.~V.\ Nazarov, Phys.\ Rev.\ Lett.\ \textbf{98}, 077003 (2007).

\bibitem{Asano2007a}
Y.\ Asano, Y.\ Tanaka, and A.~A.\ Golubov, Phys.\ Rev.\ Lett.\ \textbf{98}, 107002 (2007);
Y.\ Asano, Y.\ Sawa, Y.\ Tanaka, and A.~A.\ Golubov, Phys.\ Rev.~B \textbf{76}, 224525 (2007).

\bibitem{Yokoyama2007}
T.\ Yokoyama, Y.\ Tanaka, and A.~A.\ Golubov,
Phys.\ Rev.~B \textbf{75}, 134510  (2007).

\bibitem{Houzet07}
M.\ Houzet and A.~I.\ Buzdin,
Phys.\ Rev.~B \textbf{76}, 060504(R) (2007).
%{\it Long range triplet Josephson effect through a ferromagnetic trilayer }

\bibitem{Eschrig08}
M.\ Eschrig and T.\ L\"{o}fwander, Nature Phys.\ \textbf{4}, 138 (2008).

\bibitem{Galaktionov08}
A.~V.\ Galaktionov, M.~S.\ Kalenkov, and A.~D.\ Zaikin, Phys.\ Rev.~B \textbf{77}, 094520 (2008).
%{\it Josephson current and Andreev states in superconductor-half metal-superconductor heterostructures}

\bibitem{Halterman08}
K.\ Halterman, O.~T.\ Valls, and P.~H.\ Barsic, Phys.\ Rev.~B \textbf{77}, 174511 (2008);
%{\it Induced triplet pairing in clean s-wave superconductor/ferromagnet layered structures}
Cien-Te Wu, O.~T.\ Valls, and K.\ Halterman, \textit{ibid.} \textbf{86}, 014523 (2012).
%{\it Proximity effects and triplet correlations in ferromagnet/ferromagnet/superconductor nanostructures}

\bibitem{Grein09}
R.\ Grein, M.\ Eschrig, G.\ Metalidis, and G.\ Sch\"on,
Phys.\ Rev.\ Lett.\ \textbf{102}, 227005 (2009);
%{\it Spin-Dependent Cooper Pair Phase and Pure Spin Supercurrents in Strongly Polarized Ferromagnets }
R.\ Grein, T.\ L\"ofwander, and M.\ Eschrig,
Phys.\ Rev.~B \textbf{88}, 054502 (2013).
%{\it Inverse proximity effect and influence of disorder on triplet supercurrents in strongly spin-polarized ferromagnets}

\bibitem{Beri09}
B.\ B\'eri, J.~N.\ Kupferschmidt, C.~W.~J.\ Beenakker, and P.~W.\ Brouwer,
Phys.\ Rev.~B \textbf{79}, 024517 (2009);
%{\it Quantum limit of the triplet proximity effect in half-metal--superconductor junctions}
J.~N.\ Kupferschmidt and P.~W.\ Brouwer,
\textit{ibid.} \textbf{83}, 014512 (2011).
%{\it Andreev reflection at half-metal/superconductor interfaces with nonuniform magnetization}

\bibitem{Alidoust10}
J.\ Linder, T.\ Yokoyama, and A.\ Sudb{\o},
Phys.\ Rev.~B \textbf{79}, 054523 (2009);
%{\it Theory of superconducting and magnetic proximity effect in S/F structures with inhomogeneous magnetization textures and spin-active interfaces}
M.\ Alidoust and J.\ Linder,
\textit{ibid.} \textbf{82}, 224504 (2010).
%{\it Spin-triplet supercurrent through inhomogeneous ferromagnetic trilayers}

\bibitem{Volkov10}
A.~F.\ Volkov and K.~B.\ Efetov,
Phys.\ Rev.~B \textbf{81}, 144522 (2010).
%{\it Odd spin-triplet superconductivity in a multilayered superconductor-ferromagnet Josephson junction}

\bibitem{Trifunovic10}
L.\ Trifunovic and Z.\ Radovi\'c,
Phys.\ Rev.~B \textbf{82}, 020505 (2010).
%{\it Long-range spin-triplet proximity effect in Josephson junctions with multilayered ferromagnets}

\bibitem{Meng13}
H.\ Meng, X.\ Wu, and Z.\ Zheng,
Europhys. Lett. \textbf{104}, 37003 (2013).
%{\it Long-range triplet Josephson current modulated by the interface magnetization texture }


\bibitem{Tanaka}
Y.\ Tanaka and A.~A.\ Golubov, Phys.\ Rev.\ Lett.\ \textbf{98}, 037003 (2007).

\bibitem{Tanaka07e}
Y.\ Tanaka, A.~A.\ Golubov, S.\ Kashiwaya, and M.\ Ueda, Phys.\ Rev.\ Lett.\ \textbf{99}, 037005 (2007);
Y.\ Tanaka, M.\ Sato, and N.\ Nagaosa, J.~Phys.\ Soc.\ Jpn.\ \textbf{81}, 011013 (2012).

\bibitem{Tanuma}
Y.\ Tanaka, Y.\ Tanuma, and A.~A.\ Golubov, Phys.\ Rev.~B \textbf{76}, 054522 (2007).

%%%%%%%%%%%%%%%%%%%%%%%%%%%%%%%%%%%%%%%%%%%%%%%%%%%
% Surface Andreev Bound state
%%%%%%%%%%%%%%%%%%%%%%%%%%%%%%%%%%%%%%%%%%%%%%%%%%%%%%%%
\bibitem{Buchholtz}
L.~J.\ Buchholtz and G.\ Zwicknagl, Phys.\ Rev.~B \textbf{23} (1981) 5788.

\bibitem{Hara}
J.\ Hara and K.\ Nagai, Prog.\ Theor.\ Phys.\ \textbf{76} (1986) 1237.

\bibitem{Hu}
C.~R.\ Hu, Phys.\ Rev.\ Lett.\ \textbf{72} 1526 (1994).

\bibitem{Tanaka95}
Y.\ Tanaka and S.\ Kashiwaya, Phys.\ Rev.\ Lett.\ \textbf{74}, 3451 (1995);
Phys.\ Rev.~B \textbf{53}, 11957 (1996).

\bibitem{Kashiwaya}
S.\ Kashiwaya and Y.\ Tanaka, Rep.\ Prog.\ Phys.\ \textbf{63} 1641 (2000).

\bibitem{Suzuki}
S.-I.\ Suzuki and Y.\ Asano, Phys.\ Rev.~B \textbf{89}, 184508 (2014).

%%%%%%%%%%%%%%%%%%%%%%%%%%%%%%%%%%%%%%%%%%%%%%%%%%%%
% Odd-frequency vortex
%%%%%%%%%%%%%%%%%%%%%%%%%%%%%%%%%%%%%%%%%%%%%%%%%%%%%%%
\bibitem{Yokoyamavortex}
T.\ Yokoyama, Y.\ Tanaka, and A.~A.\ Golubov,
Phys.\ Rev.~B \textbf{78}, 012508 (2008).

\bibitem{Tanuma09}
Y.\ Tanuma, N.\ Hayashi, Y.\ Tanaka, and A.~A.\ Golubov,
Phys.\ Rev.\ Lett.\ \textbf{102}, 117003 (2009).

\bibitem{Yokoyama10}
T.\ Yokoyama, M.\ Ichioka, and Y.\ Tanaka,
J.~Phys.\ Soc.\ Jpn.\ \textbf{79}, 034702 (2010).

%%%%%%%%%%%%%%%%%%%%%%%%%%%%%%%%%%%%%%%%%%%%%%%%%%%%%%
% Odd-frequency Majorana and topological superconductor
%%%%%%%%%%%%%%%%%%%%%%%%%%%%%%%%%%%%%%%%%%%%%%%%%%%
\bibitem{Asano2013}
Y.\ Asano and Y.\ Tanaka, Phys.\ Rev.~B \textbf{87},
104513 (2013).

\bibitem{Stanev}
V.\ Stanev and V.\ Galitski,
Phys.\ Rev.~B \textbf{89}, 174521 (2014).

\bibitem{Wakatsuki}
R.\ Wakatsuki, M.\ Ezawa, Y.\ Tanaka, and N.\ Nagaosa,
Phys.\ Rev.~B \textbf{90}, 014505 (2014).

\bibitem{Balatsky1}
A.~M.\ Black-Schaffer and A.~V.\ Balatsky,
Phys.\ Rev.~B \textbf{87}, 220506 (2013).

\bibitem{Ebisu}
H.\ Ebisu, K.\ Yada, H.\ Kasai, and Y.\ Tanaka,
Phys.\ Rev.~B \textbf{91}, 054518 (2015).


\bibitem{Vorontsov}
A.~B.\ Vorontsov, I.\ Vekhter, and M.\ Eschrig, Phys.\ Rev.\ Lett.\ \textbf{101}, 127003 (2008).

\bibitem{GrMiEs}
R.\ Grein, J.\ Michelsen, and M.\ Eschrig, J.~Phys.: Conf.\ Ser.\ \textbf{391}, 012149 (2012).



%%%%%%%%%%%%%%%%%%%%%%%%%%%%%%%%%%%%%%%%%%%%%%%%%%
% Odd-frequency two band systems
%%%%%%%%%%%%%%%%%%%%%%%%%%%%%%%%%%%%%%%%%%%%%%%%%%%%%%%%%%%%
\bibitem{Balatsky2}
A.~M.\ Black-Schaffer and A.~V.\ Balatsky,
Phys.\ Rev.~B \textbf{88}, 104514 (2013).



\bibitem{Tanaka2004}
Y.\ Tanaka and S.\ Kashiwaya, Phys.\ Rev.~B \textbf{70}, 012507 (2004).

\bibitem{Tanaka2006}
Y.\ Asano, Y.\ Tanaka, and S.\ Kashiwaya, Phys.\ Rev.\ Lett.\ \textbf{96}, 097007 (2006).

\bibitem{Asano2007}
Y.\ Asano, Y.\ Tanaka, A.~A.\ Golubov, and S.\ Kashiwaya, Phys.\ Rev.\ Lett.\ \textbf{99}, 067005 (2007).

\bibitem{Fominov}
Ya.~V.\ Fominov, Pis'ma Zh.\ Eksp.\ Teor.\ Fiz.\ \textbf{86}, 842 (2007)  [JETP Lett.\ \textbf{86}, 732 (2007)].

\bibitem{Linder09}
J.\ Linder, T.\ Yokoyama, A.\ Sudb{\o}, and M.\ Eschrig,
Phys.\ Rev.\ Lett.\ \textbf{102}, 107008 (2009);
J.\ Linder, A.\ Sudb{\o}, T.\ Yokoyama, R.\ Grein, and M.\ Eschrig, Phys.\ Rev.~B \textbf{81}, 214504 (2010).
%{\it Pairing Symmetry Conversion by Spin-Active Interfaces in Magnetic Normal-Metal??uperconductor Junctions}

\bibitem{Eschrig09}
M.\ Eschrig, Phys.\ Rev.~B \textbf{80}, 134511 (2009).
%{\it Scattering problem in non-equilibrium quasiclassical theory of metals and superconductors: general boundary conditions and applications}

%%%%%%%%%%%%%%%%%%%%%%%%%%%%%%%%%%%%%%%%%%%%%%%%%%%%%%%
\bibitem{Asano}
Y.\ Asano, A.~A.\ Golubov, Ya.~V.\ Fominov, Y.\ Tanaka, Phys.\ Rev.\ Lett.\ \textbf{107}, 087001 (2011).
%%%%%%%%%%%%%%%%%%%%%%%%%%%%%%%%%%%%%%%%%%%%%%%%%%%%%%%%%%
% Anomalous Proximity including negative \rho
%%%%%%%%%%%%%%%%%%%%%%%%%%%%%%%%%%%%%%%%%%%%%%%%%%%%

\bibitem{Tanaka2005}
Y.\ Tanaka, Y.\ Asano, A.~A.\ Golubov, and S.\ Kashiwaya, Phys.\ Rev.~B \textbf{72}, 140503(R) (2005).

\bibitem{Yokoyama2011}
T.\ Yokoyama, Y.\ Tanaka, and N.\ Nagaosa, Phys.\ Rev.\ Lett.\ \textbf{106}, 246601 (2011).

\bibitem{Higashitani}
S.\ Higashitani, H.\ Takeuchi, S.\ Matsuo, Y.\ Nagato, and K.\ Nagai, Phys.\ Rev.\ Lett.\ \textbf{110}, 175301 (2013).

\bibitem{Alidoust}
M.\ Alidoust, K.\ Halterman, and J.\ Linder, Phys.\ Rev.~B \textbf{89}, 054508 (2014).


\bibitem{HeidBazaliy}
R.\ Heid, Ya.~B.\ Bazaliy, V.\ Martisovits, and D.~L.\ Cox, Phys.\ Rev.\ Lett.\ \textbf{74}, 2571 (1995).

\bibitem{CoxZawadowski}
D.~L.\ Cox and A.\ Zawadowski, Adv.\ Phys.\ \textbf{47}, 599 (1998).

\bibitem{Mironov}
S.\ Mironov, A.\ Mel'nikov, and A.\ Buzdin, Phys.\ Rev.\ Lett.\ \textbf{109}, 237002 (2012).

\bibitem{Higashitani97}
S.\ Higashitani,
J.~Phys.\ Soc.\ Jpn.\ \textbf{66},
2556 (1997).

\bibitem{Barash}
Yu.~S.\ Barash, M.~S.\ Kalenkov, and J.\ Kurkij{\"a}rvi,
Phys.\ Rev.~B \textbf{62}, 6665 (2000).

%%%%%%%%%%%%%%%%%%%%%%%%%%%%%%%%%%%%%%%%%%%%%%%%%%%%%%%%%%%%%
%  Ferromagnet experiments
%%%%%%%%%%%%%%%%%%%%%%%%%%%%%%%%%%%%%%%%%%%%%%%%%%%%%%%%%%%%%

\bibitem{Keizer}
R.~S.\ Keizer, S.~T.~B.\ Goennenwein, T.~M.\ Klapwijk, G.\ Miao, G.\ Xiao, and A.\ Gupta, Nature (London)
\textbf{439}, 825 (2006).

\bibitem{Petrashov06}
I.\ Sosnin, H.\ Cho, V.~T.\ Petrashov, A.~F.\ Volkov,
Phys.\ Rev.\ Lett.\ \textbf{96}, 157002 (2006).
%{\it Superconducting phase coherent electron transport in proximity conical ferromagnets}

\bibitem{Anwar}
M.~S.\ Anwar, F.\ Czeschka, M.\ Hesselberth, M.\ Porcu, and J.\ Aarts, Phys.\ Rev.~B \textbf{82}, 100501 (2010);
M.~S.\ Anwar and J.\ Aarts,
Supercond.\ Sci.\ Technol.\ \textbf{24}, 024016 (2011);
M.~S.\ Anwar, M.\ Veldhorst, A.\ Brinkman, and J.\ Aarts,
Appl.\ Phys.\ Lett.\ \textbf{100}, 052602 (2012).

\bibitem{Khaire10}
T.~S.\ Khaire, M.~A.\ Khasawneh, W.~P.\ Pratt,~Jr., and N.~O.\ Birge,
Phys.\ Rev.\ Lett.\ \textbf{104}, 137002 (2010).
%{\it Observation of Spin-Triplet Superconductivity in Co-Based Josephson Junctions}

\bibitem{Robinson10}
J.~W.~A.\ Robinson, J.~D.~S.\ Witt, and M.~G.\ Blamire,
Science \textbf{329}, 59 (2010);
%{\it Controlled Injection of Spin-Triplet Supercurrents into a Strong Ferromagnet}
J.~D.~S.\ Witt, J.~W.~A.\ Robinson, and M.~G.\ Blamire,
Phys.\ Rev.~B \textbf{85}, 184526 (2012).
%{\it Josephson junctions incorporating a conical magnetic holmium interlayer}

\bibitem{Sprungmann10}
D.\ Sprungmann, K.\ Westerholt, H.\ Zabel, M.\ Weides, and H.\ Kohlstedt,
Phys.\ Rev.~B \textbf{82}, 060505(R) (2010).
%{\it Evidence for triplet superconductivity in Josephson junctions with barriers of the ferromagnetic Heusler alloy Cu$_2$MnAl}

\bibitem{Lofwander10}
T.\ L\"ofwander, R.\ Grein, and M.\ Eschrig,
Phys.\ Rev.\ Lett.\ \textbf{105}, 207001 (2010).
%{\it Is CrO$_2$ Fully Spin Polarized? Analysis of Andreev Spectra and Excess Current}

\bibitem{Gu10}
J.~Y.\ Gu, J.\ Kusnadi, and Ch.-Y.\ You,
Phys. Rev. B \textbf{81}, 214435 (2010); \textbf{82}, 099905(E) (2010).
%{\it Proximity effect in a superconductor/exchange-spring-magnet hybrid system}

\bibitem{Boden}
K.~M.\ Boden, W.~P.\ Pratt~Jr., and N.~O.\ Birge, Phys.\ Rev.~B \textbf{84}, 020510 (2011).

\bibitem{Kalcheim12}
Y.\ Kalcheim, O.\ Millo, M.\ Egilmez, J.~W.~A.\ Robinson, and M.~G.\ Blamire,
Phys.\ Rev.~B \textbf{85}, 104504 (2012).
%{\it Evidence for anisotropic triplet superconductor order parameter in half-metallic ferromagnetic La$_{0.7}$Ca$_{0.3}$Mn$_3$O proximity coupled to superconducting Pr$_{1.85}$Ce$_{0.15}$CuO$_4$}

\bibitem{Visani12}
C.\ Visani, Z.\ Sefrioui, J.\ Tornos, C.\ Leon, J.\ Briatico, M.\ Bibes, A.\ Barthelemy, J.\ Santamaria, and J.~E.\ Villegas,
Nature Phys.\ \textbf{8}, 539 (2012).
%{\it Equal-spin Andreev reflection and long-range coherent transport in high-temperature superconductor/half-metallic ferromagnet junctions}

\bibitem{Yates13}
K.~A.\ Yates, M.~S.\ Anwar, J.\ Aarts, O.\ Conde, M.\ Eschrig, T.\ L\"ofwander, L.~F.\ Cohen, Europhys.\ Lett.\ \textbf{103}, 67005 (2013).
%{\it Andreev spectroscopy of CrO$_2$ thin films on TiO$_2$ and Al$_2$O$_3$}

\bibitem{Piano}
S.\ Piano, R.\ Grein, C.~J.\ Mellor, K.\ V\'yborn\'y, R.\ Campion, M.\ Wang, M.\ Eschrig, and B.~L.\ Gallagher, Phys.\ Rev.~B \textbf{83}, 081305(R) (2011).

\bibitem{HublerWolf}
F.\ H\"ubler, M.~J.\ Wolf, T.\ Scherer, D.\ Wang, D.\ Beckmann, and H.~v.\ L\"ohneysen, Phys.\ Rev.\ Lett.\ \textbf{109}, 087004 (2012).

\bibitem{Flokstra}
M.~G.\ Flokstra, T.~C.\ Cunningham, J.\ Kim, N.\ Satchell, G.\ Burnell, P.~J.\ Curran, S.~J.\ Bending, C.~J.\ Kinane, J.~F.~K.\ Cooper, S.\ Langridge, A.\ Isidori, N.\ Pugach, M.\ Eschrig, and S.~L.\ Lee, Phys.\ Rev.~B \textbf{91}, 060501(R) (2015).




\bibitem{Solenov}
D.\ Solenov, I.\ Martin, and D.\ Mozyrsky, Phys.\ Rev.~B \textbf{79}, 132502 (2009).

\bibitem{Kusunose}
H.\ Kusunose, Y.\ Fuseya, and K.\ Miyake, J.~Phys.\ Soc.\ Jpn.\ \textbf{80}, 054702 (2011).
% On the Puzzle of Odd-Frequency Superconductivity


\bibitem{BCScomm}
The conventional BCS case \cite{AGD} corresponds to the singlet interaction with $V_s(1-2) = \lambda \delta(1-2)$, which in turn corresponds to the second-quantized interaction Hamiltonian $\hat H_\mathrm{int} = (\lambda/2) \int d\mathbf r\, \hat\psi_\alpha^\dagger (\mathbf r) \hat\psi_\beta^\dagger (\mathbf r) \hat\psi_\beta (\mathbf r) \hat\psi_\alpha (\mathbf r)$.

\bibitem{AGD}
A.~A.\ Abrikosov, L.~P.\ Gorkov, and I.~E.\ Dzyaloshinski, \textit{Methods of Quantum Field Theory in Statistical Physics} (Dover, New York, 1975).

\bibitem{Higashitani09}
S.\ Higashitani, Y.\ Nagato, and K.\ Nagai,
J.~Low Temp.\ Phys.\ \textbf{155}, 83 (2009).


\bibitem{Nazarov}
Yu.~V.\ Nazarov, Superlattices Microstruct.\ \textbf{25}, 1221 (1999);
Y.\ Tanaka, Yu.~V.\ Nazarov, and S.\ Kashiwaya, Phys.\ Rev.\ Lett.\ \textbf{90}, 167003  (2003).

\bibitem{KL}
M.~Yu.\ Kuprianov and V.~F.\ Lukichev, Zh.\ Eksp.\ Teor.\ Fiz.\ \textbf{94}, 139 (1988) [Sov.\ Phys.\ JETP \textbf{67}, 1163 (1988)].

\bibitem{Kusunose12}
%A possibility of a coexisting order parameter has been discussed
%in the absence of time-reversal symmetry by
H.\ Kusunose, M.\ Matsumoto, and M.\ Koga, Phys.\ Rev.~B \textbf{85}, 174528 (2012);
M.\ Matsumoto, M.\ Koga, and H.\ Kusunose, J.~Phys.\ Soc.\ Jpn.\ \textbf{81}, 033702 (2012).

\bibitem{Asano2014}
Y.\ Asano, Ya.~V.\ Fominov, and Y.\ Tanaka, Phys.\ Rev.~B \textbf{90}, 094512 (2014).

\end{thebibliography}
\end{document}